\documentclass[preprint,3p,times]{elsarticle}

\usepackage{amsthm}
\usepackage{amssymb}
\usepackage{mathtools}

\usepackage{float}
\floatstyle{plaintop}
\restylefloat{table}

\newcommand\Tstrut{\rule{0pt}{2.6ex}}         
\newcommand\Bstrut{\rule[-0.9ex]{0pt}{0pt}}   

\usepackage{caption}

\usepackage{color}
\usepackage{soul}
\usepackage{cancel}
\definecolor{myc}{rgb}{1,1,1}

\usepackage{algorithm,algorithmicx}
\usepackage[noend]{algpseudocode}
\usepackage{varioref}

\newcommand\hcancel[2][black]{\setbox0=\hbox{$#2$}%
\rlap{\raisebox{.45\ht0}{\textcolor{#1}{\rule{\wd0}{1pt}}}}#2} 

\journal{Journal of Computational Physics}

\begin{document}

\begin{frontmatter}

\title{A fourth-order accurate compact difference scheme for solving the three-dimensional Poisson equation with arbitrary boundaries}

\makeatletter
\makeatother

\author{Shirzad Hosseinverdi\corref{cor1}}
\cortext[cor1]{Corresponding author}
\ead{shirzadh@email.arizona.edu}

\author{Hermann F. Fasel}
\address{Department of Aerospace and Mechanical Engineering, University of Arizona, Tucson, AZ 85721, USA}

\begin{abstract}
This paper presents an efficient high-order sharp-interface method for solving the three-dimensional (3D) Poisson equation with Dirichlet boundary conditions on a nonuniform Cartesian grid with irregular domain boundaries. The new approach is based on the combination of the fourth-order compact finite difference scheme and the preconditioned stabilized biconjugate-gradient (BiCGSTAB) method. Contrary to the original immersed interface method by LeVeque \& Li \citep{17}, the new method does not require jump corrections, instead, the (regular) compact finite difference stencil is adjusted at the irregular grid points (in the vicinity of the interfaces of the immersed bodies) to obtain a solution that is sharp across the interface while keeping the fourth-order global accuracy. The contribution of the present work is the design of a fourth-order-accurate 3D Poisson solver whose accuracy and efficiency does not deteriorate in the presence of an immersed boundary. This is attributed to (i) the modification of the discrete operators near immersed boundaries does not lead to a wide grid stencil due to the compact nature of the discretization and (ii) a preconditioning technique whose efficiency and cost are independent of the complexity of the geometry and the presence or not of an immersed boundary. The accuracy and computational efficiency of the proposed algorithm is demonstrated and validated over a range of problems including smooth and irregular boundaries. The test cases show that the new method is fourth-order accurate in the maximum norm whether an immersed boundary is present or not, on uniform or nonuniform grids. Furthermore, the efficiency of the preconditioned BiCGSTAB is demonstrated with regard to convergence rate and ``extra'' floating-point operation ($FLOP_{extra}$) which is due to the presence of immersed boundaries. It is shown that the solution method is equally efficient for domains with and without irregular boundaries, with a negligible $FLOP_{extra}$ in the presence of immersed boundaries.
\end{abstract}


\end{frontmatter}

\section{Introduction} \label{sec:intro}
In this paper, a new numerical method is presented for solving the three-dimensional (3D) Poisson equation with Dirichlet boundary conditions in the 3D rectangular box $\Omega$ which can contain an arbitrary immersed body (IB) with boundary $\Gamma$,
\begin{equation}
\begin{aligned}
\bigg(\frac{\partial^2}{\partial x^2} + \frac{\partial^2}{\partial y^2} + \frac{\partial^2}{\partial z^2}\bigg)~u(x,y,z)&=f(x,y,z),  \quad (x,y,z)\in \Omega ,\\
u(x,y,z)&=g(x,y,z), \quad (x,y,z)\in \partial\Omega~\&~ \Gamma,
\label{eq:poisson}
\end{aligned}
\end{equation}
where $u(x,y,z)$ is an unknown function defined in the domain $\Omega$ with prescribed boundary values $g(x,y,z)$ on its boundary $\partial\Omega$ and $f(x,y,z)$ is a known source function. There has been a great deal of research work on the development of high-order finite difference numerical solution of 3D Poisson equations in the past two decades \citep{1,2,3,4,5}, with particular focus on the fourth-order accurate compact difference scheme. However, the majority of these schemes are limited to simple domains with uniform mesh distributions. For many practical applications, however, the numerical computation of the 3D Poisson equation in a domain with irregular boundaries is required. In the presence of an IB, the domain $\Omega$ is divided by the surface $\Gamma$ into two subdomains $\Omega^{+}$ and $\Omega^{-}$ corresponding to the outside and inside of the immersed body, respectively (see Fig. \ref{fig:grid_point_types}). One would typically solve Eq. (\ref{eq:poisson}) defined on the open region $\Omega^{+}$ with boundary conditions on $\partial\Omega$ (the outer boundary which conforms to the computational domain), and $\Gamma$, the immersed boundary. The solution in the region $\Omega^{-}$ may or may not be of interest. In either case, the immersed boundary $\Gamma$ represents a singularity, thus field variables and/or their derivatives will be discontinuous across the immersed boundary. In the present work, we refer to $\Omega^{-}$ and $\Omega^{+}$ solid and fluid regions, respectively, and the solution inside the immersed body, $\Omega^{-}$, is trivial and set to zero.

Most of the numerical algorithms capable of handling complex geometries use body-fitted structured or unstructured grids. However, generating high-quality structured grids is generally cumbersome and it becomes very laborious as the complexity of the geometry increases. The grid generation process becomes a very difficult task, even for the simplest geometries, when more than one body is located within the domain. Grids with poor qualities (smoothness, orthogonality, aspect ratio) could negatively impact the accuracy and convergence properties of the numerical method for solving the Poisson equation. On the other hand, unstructured grids which are more suited for complex geometries, suffer from slow rate of convergence which could lead to a substantial increase in the computation time \citep{6}. Compared to structured grids, unstructured grids require a larger amount of memory and the extension of the numerical schemes to higher-order is not straightforward.

An alternative approach is to discretize Eq. (\ref{eq:poisson}) on a fixed Cartesian grid which is allowed to intersect with an immersed boundary $\Gamma$. As a result, the grid does not conform to the solid boundaries. The main advantage in this case lies in the use of Cartesian grids for which there is almost no grid generation cost and high-order numerical methods developed for Cartesian grids, can be employed. A main challenge here is the imposition of boundary conditions on the immersed boundaries such that the accuracy and efficiency of the Cartesian solvers is maintained. Several methods have been proposed in the past to handle the singularity associated with the immersed boundary. Generally, they can be classified into two categories: the so-called immersed boundary methods (IBM), and immersed interface methods (IIM). 

The original IBM was pioneered by Peskin \citep{7,8} to handle elastic boundaries for simulating blood flow in the heart. In Peskin`s approach, the boundary conditions are enforced through a smooth forcing term added to the Navier-Stokes equations. This type of IBM is classified as continuous forcing (diffuse) approach. One disadvantage of diffuse methods is that the effect of the boundary is distributed over a band of several grid points which causes a smearing effect near the boundary. This smearing has a detrimental effect on the accuracy of the numerical scheme. The other approach in IBM is discrete forcing (sharp) methods \citep{9}. In the discrete IBM, the numerical discretization near the immersed boundary is modified such as to account directly for the presence of the boundary, so that the interface remains ``sharp``. The ``ghost cell`` in the finite-difference method and the ``cut-cell`` within the finite volume framework fall into this category. 

In the ghost cell method, boundary conditions at the immersed boundary are enforced through ghost cells. The ghost cell refers to a fictitious cell that lies inside the solid region and whose value is extrapolated from the boundary condition at the immersed boundary and the surrounding fluid points (in the fluid region). While the approach is well suited for achieving second-order accuracy, extension to higher-order formulations is problematic. Higher-order formulations require large interpolation stencils which could lead to robustness issues. For example, for a fourth-order accurate scheme in three dimensions, at least 35 points are required to evaluate the ghost cell \citep{10}. Hence, most of the existing immersed-boundary ghost cell methods are of second-order accuracy \citep{11,12,13}. On the other hand, being based on the finite volume approach, the cut-cell method is designed to provide better conservation properties, especially for cells at the immersed boundary \citep{14,15}. Grid cells cut by the immersed boundary, whose cell centers are inside the fluid region, are reshaped by discarding the portion of these cells that lies in the solid. The disadvantages of this method are related with the process of cutting the cells. The reshaping may in some cases result in very small grid cells with an adverse impact on numerical stability. Hence, the cut cells near boundaries must be adjusted, modified and/or merged. In case of very complex geometries, this process may even fail to preserve the geometrical representation of bodies. The extension of this approach to 3D problems becomes very complicated and a non-trivial task. Similar to the ghost cell method, most of the cut-cell based IMB methods are second-order accurate such as the one adopted by Seo and Mittal \citep{16} to solve the pressure Poisson equation as part of the solution of Navier-Stokes equations.

Standard finite-difference schemes fail when applied to non-smooth functions because the underlying Taylor expansions upon which they are based are invalid. To avoid this problem, jump correction terms need to be added to the finite difference schemes at jump discontinuities for the function value and its derivatives. Based on this key idea, LeVeque \& Li \cite{17} developed a second-order accurate sharp immersed interface method to solve elliptic problems with discontinuous and non-smooth solutions. The advantages of this technique are that boundary conditions are imposed directly at the location of the boundary and high-order local accuracy can be achieved around the immersed boundary. An important application of sharp IIM is the treatment of problems defined in an irregular domain where the solution inside or the outside the interface is not of interest and is trivial \citep{18,20}. Apart from the IBM and IIM, other approaches have been proposed in the literature to solve 3D elliptic equations with irregular boundaries such as the second-order accurate Shortley–Weller embedded finite-difference method for the solution of the 3D Poisson equation \citep{21} and high-order matched interface and boundary method for solving elliptic equations with discontinuous coefficients and non-smooth interfaces \citep{22}.

Eq. (\ref{eq:poisson}) is an elliptic partial differential equation with a broad range of applications in electromagnetism, geophysics, astrophysics and fluid mechanics. Therefore, there is great interest to develop highly-accurate and computationally efficient numerical methods for the numerical solution of Eq. (\ref{eq:poisson}) for simple and complex geometries. Building on our previous research \citep{HOSSEINVERDI2018912}, the main goal of this paper is to present a solver for the 3D Poisson equation in a domain with immersed boundaries on a nonuniform grid which combines high accuracy and high efficiency.  The objective of the present work is twofold: (1) Developing a uniformly fourth-order-accurate finite-difference algorithm on an irregularly shaped boundary, and (2) designing an efficient and cost-effective iterative solver which easily and efficiently accommodates the irregular immersed boundaries. The paper is organized as follows: The discretization of Eq. (\ref{eq:poisson}) on a nonuniform grid is presented in Section \ref{sec:discr}. It contains the numerical procedure to construct a fourth-order accurate compact difference stencils for regular and irregular grid points, as well as a formal proof for the order of accuracy. A spectral analysis of the resulting coefficient matrices for simple and irregular domains and the solution strategy of the discretized equations is explained in Section \ref{sec:sol}. Then, in Section \ref{sec:results}, the proposed method is validated for several test cases, which demonstrate the high efficiency and confirm the fourth-order convergence. A summary and conclusions are provided in Section \ref{sec:end}.

\section{Discretization} \label{sec:discr}
\begin{figure}[htbp!]
    \centerline{    
    \includegraphics[width=0.75\textwidth]{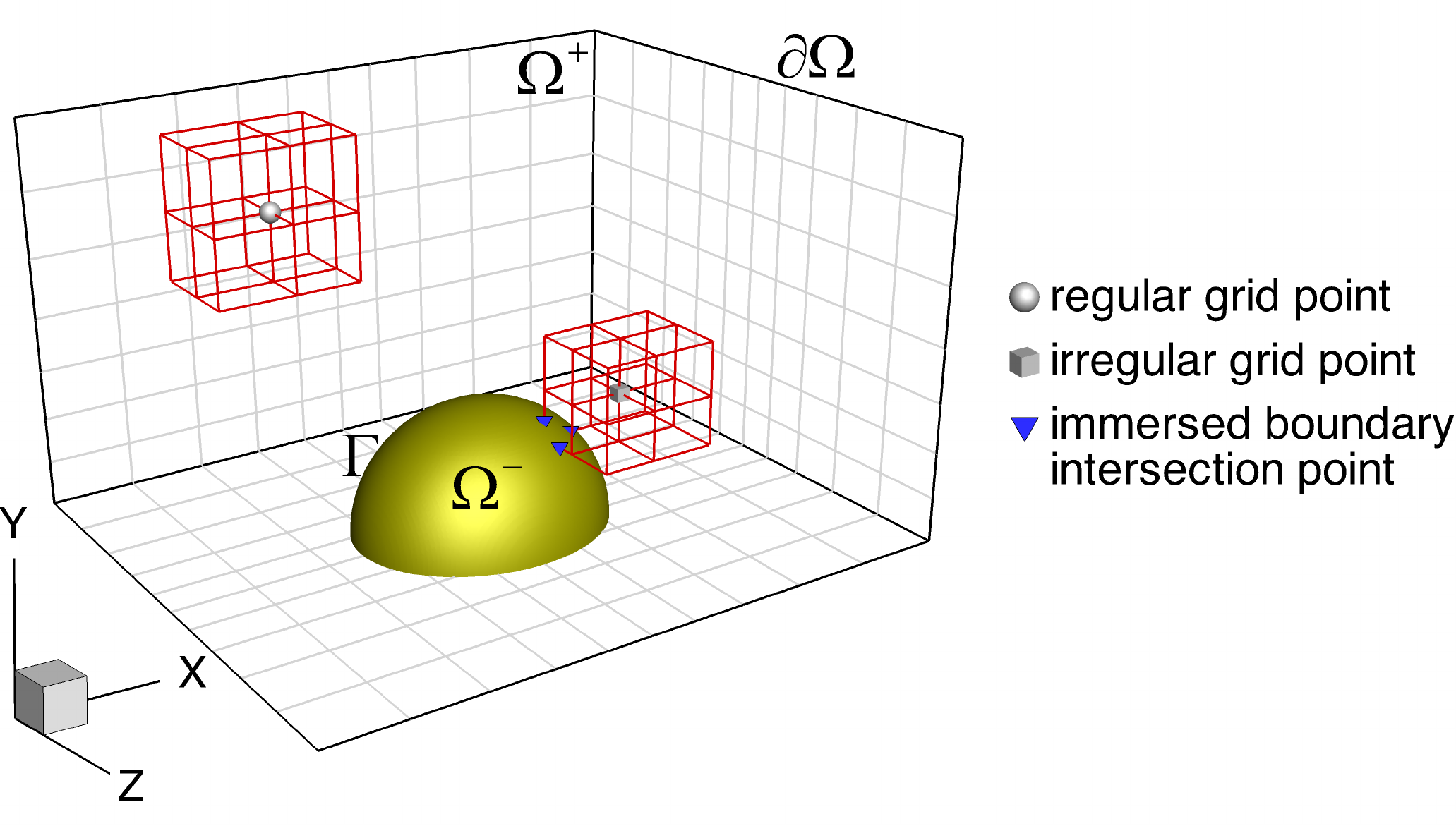}}
    \caption{Representation of the different type of grid points.}
    \label{fig:grid_point_types}
\end{figure}

Eq. (\ref{eq:poisson}) is solved in a cubic domain $\Omega$ defined on $[a_1,a_2]\times[b_1,b_2]\times[c_1,c_2]$. The domain is divided into $nx\times ny\times nz$ uniform/nonuniform cells by the points $a_1=x_1<x_2<\cdots<x_{nx}=a_2$, $b_1=y_1<y_2<\cdots<y_{ny}=b_2$ and $c_1=z_1<z_2<\cdots<z_{nz}=c_2$. The discretization of Eq. (\ref{eq:poisson}) for regular and irregular grid points are discussed in detail in this section. A grid point is said to be regular if all the 26 neighboring grid points are outside the immersed body, otherwise, it is defined as an irregular point. A grid point is called a solid point if it lies inside an immersed body. Furthermore, the locations where the immersed/irregular boundary intersects with the grid are called immersed/irregular boundary intersection (IBI) points. The IBI points are the locations where the boundary conditions can be enforced. Regular, irregular and IBI points are illustrated in Fig. \ref{fig:grid_point_types}.

\subsection{Fourth-order compact difference scheme on nonuniform grids} \label{sec:cfd_reg}
In this section, the discretization for regular grid points away from the immersed boundary is discussed, so that the compact difference scheme is well-defined and valid. The discretization of Eq. (\ref{eq:poisson}) is based on three one-dimensional, fourth-order compact finite-difference schemes for second derivatives in $x$, $y$ and $z$:
\begin{equation}
\begin{aligned}
(L_{xx}u_{xx})_{j,k} &=(R_{xx}u)_{j,k}~, \\
(L_{yy}u_{yy})_{i,k} &=(R_{yy}u)_{i,k}~, \\
(L_{zz}u_{zz})_{i,j} &=(R_{zz}u)_{i,j}~,
\end{aligned}
\label{eq:cfd_xxyyzz}
\end{equation}
where the finite difference (FD) operators are given by
\begin{equation}
\begin{alignedat}{2}
(L_{xx}u_{xx})_{j,k} &= a^x_i~u_{{xx}_{i-1,j,k}} &+~ b^x_i~u_{{xx}_{i,j,k}} &+ c^x_i~u_{{xx}_{i+1,j,k}}~, \\
(L_{yy}u_{yy})_{i,k} &= a^y_j~u_{{yy}_{i,j-1,k}} &+~ b^y_j~u_{{yy}_{i,j,k}} &+ c^y_j~u_{{yy}_{i,j+1,k}}~, \\
(L_{zz}u_{zz})_{i,j} &= a^z_k~u_{{zz}_{i,j,k-1}} &+~ b^z_k~u_{{zz}_{i,j,k}} &+ c^z_k~u_{{zz}_{i,j,k+1}}~,
\end{alignedat}
\label{eq:cfd_lhs_xxyyzz}
\end{equation}
and
\begin{equation}
\begin{aligned}
(R_{xx}u)_{j,k} &= ar^x_i~u_{i-1,j,k} + br^x_i~u_{i,j,k} + cr^x_i~u_{i+1,j,k}~, \\
(R_{yy}u)_{i,k} &= ar^y_j~u_{i,j-1,k} + br^y_j~u_{i,j,k} + cr^y_j~u_{i,j+1,k}~, \\
(R_{zz}u)_{i,j} &= ar^z_k~u_{i,j,k-1} + br^z_k~u_{i,j,k} + cr^z_k~u_{i,j,k+1}~. 
\end{aligned}
\label{eq:cfd_rhs_xxyyzz}
\end{equation}

Here $u_{xx}$, $u_{yy}$ and $u_{zz}$ represent numerical approximations to the second partial derivatives in $x$, $y$ and $z$ directions, respectively. Coefficients for compact FD operators in the $x-$direction in Eqs. (\ref{eq:cfd_lhs_xxyyzz}) and (\ref{eq:cfd_rhs_xxyyzz}), obtained by matching the coefficients in the Taylor expansion about $u_{i,j,j}$ in the $x-$direction, are given by
\begin{equation}
\begin{alignedat}{3}
a^x_{i} &= \frac {dx_{f}~(dx_{b}^{2}+dx_{b}dx_{f}-dx_{f}^{2})} {dx_{f}^{3}+4dx_{f}^{2}dx_{b}+4dx_{f}dx_{b}^{2}+dx_{b}^{3}}~, ~~~~~
&ar^x_{i} &= \frac {12dx_{f}} {dx_{f}^{3}+4dx_{f}^{2}dx_{b}+4dx_{f}dx_{b}^{2}+dx_{b}^{3}}~,\\
b^x_{i} &= 1~, ~~~~~
&br^x_{i} &= \frac {-12} {(dx_{f}^{2}+3dx_{f}dx_{b}+dx_{b}^{2})}~,\\
c^x_{i} &= \frac {dx_{b}~(dx_{f}^{2}+dx_{b}dx_{f}-dx_{b}^{2})} {dx_{f}^{3}+4dx_{f}^{2}dx_{b}+4dx_{f}dx_{b}^{2}+dx_{b}^{3}}~, ~~~~~
&cr^x_{i} &= \frac {12dx_{b}} {dx_{f}^{3}+4dx_{f}^{2}dx_{b}+4dx_{f}dx_{b}^{2}+dx_{b}^{3}}~,
\end{alignedat}
\label{eq:app_xxx}
\end{equation}
where $dx_f=x_{i+1}-x_i$ and $dx_b=x_i-x_{i-1}$. The coefficients for compact FD operators in the $y-$ and $z-$directions in Eqs. (\ref{eq:cfd_lhs_xxyyzz}) and (\ref{eq:cfd_rhs_xxyyzz}) are the same except that $x$ is replaced with $y$ and $z$, respectively. Combining Eq. (\ref{eq:cfd_xxyyzz}) at three consecutive $x-$, $y-$ and $z-$locations centered at point ($i,j,k$) leads to
\begin{equation}
 \label{eq:lhsrhs_xxyyzz}
\begin{alignedat}{3} 
a^y_j~a^z_k~(L_{xx}u_{xx}-R_{xx}u)_{j-1,k-1} &+ b^y_j~a^z_k~(L_{xx}u_{xx}-R_{xx}u)_{j  ,k-1} &{}+{}& c^y_j~a^z_k~(L_{xx}u_{xx}-R_{xx}u)_{j+1,k-1} &{}+{}& \\
a^y_j~b^z_k~(L_{xx}u_{xx}-R_{xx}u)_{j-1,k\textcolor{myc}{+1}} &+ b^y_j~b^z_k~(L_{xx}u_{xx}-R_{xx}u)_{j  ,k\textcolor{myc}{+1}} &{}+{}& c^y_j~b^z_k~(L_{xx}u_{xx}-R_{xx}u)_{j+1,k\textcolor{myc}{+1}} &{}+{}&  \\
a^y_j~c^z_k~(L_{xx}u_{xx}-R_{xx}u)_{j-1,k+1} &+ b^y_j~c^z_k~(L_{xx}u_{xx}-R_{xx}u)_{j  ,k+1} &{}+{}& c^y_j~c^z_k~(L_{xx}u_{xx}-R_{xx}u)_{j+1,k+1} &{}+{}& \\
a^x_i~a^z_k~(L_{yy}u_{yy}-R_{yy}u)_{i-1,k-1} &+ b^x_i~a^z_k~(L_{yy}u_{yy}-R_{yy}u)_{i  ,k-1} &{}+{}& c^x_i~a^z_k~(L_{yy}u_{yy}-R_{yy}u)_{i+1,k-1} &{}+{}& \\
a^x_i~b^z_k~(L_{yy}u_{yy}-R_{yy}u)_{i-1,k\textcolor{myc}{+1}} &+ b^x_i~b^z_k~(L_{yy}u_{yy}-R_{yy}u)_{i  ,k\textcolor{myc}{+1}} &{}+{}& c^x_i~b^z_k~(L_{yy}u_{yy}-R_{yy}u)_{i+1,k\textcolor{myc}{+1}} &{}+{}& \\
a^x_i~c^z_k~(L_{yy}u_{yy}-R_{yy}u)_{i-1,k+1} &+ b^x_i~c^z_k~(L_{yy}u_{yy}-R_{yy}u)_{i  ,k+1} &{}+{}& c^x_i~c^z_k~(L_{yy}u_{yy}-R_{yy}u)_{i+1,k+1} &{}+{}& \\
a^x_i~a^y_j~(L_{zz}u_{zz}-R_{zz}u)_{i-1,j-1~} &+ b^x_i~a^y_j~(L_{zz}u_{zz}-R_{zz}u)_{i  ,j-1~} &{}+{}& c^x_i~a^y_j~(L_{zz}u_{zz}-R_{zz}u)_{i+1,j-1~} &{}+{}& \\
a^x_i~b^y_j~(L_{zz}u_{zz}-R_{zz}u)_{i-1,j~~~~} &+ b^x_i~b^y_j~(L_{zz}u_{zz}-R_{zz}u)_{i  ,j~~~~} &{}+{}& c^x_i~b^y_j~(L_{zz}u_{zz}-R_{zz}u)_{i+1,j~~~~} &{}+{}& \\
a^x_i~c^y_j~(L_{zz}u_{zz}-R_{zz}u)_{i-1,j+1~} &+ b^x_i~c^y_j~(L_{zz}u_{zz}-R_{zz}u)_{i  ,j+1~} &{}+{}& c^x_i~c^y_j~(L_{zz}u_{zz}-R_{zz}u)_{i+1,j+1~} &{}={}&~0.
\end{alignedat}
\end{equation}

Applying the FD operators in Eq. (\ref{eq:lhsrhs_xxyyzz}) and using Eq. (\ref{eq:poisson}) leads to a 27-points, fourth-order compact scheme at the regular grid point ($i,j,k$) inside the computational domain as follows:
\begin{equation}
 \label{eq:A27}
  \begin{alignedat}{4}
A_{i,j,k}^{1\textcolor{myc}{1}}~u_{i-1,j+1,k-1}  &~+~ 
A_{i,j,k}^{2\textcolor{myc}{1}}~u_{i  ,j+1,k-1}  &~+~ 
A_{i,j,k}^{3\textcolor{myc}{1}}~u_{i+1,j+1,k-1}  &~+  \\  
A_{i,j,k}^{4\textcolor{myc}{1}}~u_{i-1,j\textcolor{myc}{+1},k-1}  &~+~ A_{i,j,k}^{5\textcolor{myc}{1}}~u_{i  ,j\textcolor{myc}{+1},k-1}  &~+~ A_{i,j,k}^{6\textcolor{myc}{1}}~u_{i+1,j\textcolor{myc}{+1},k-1}  &~+  \\  
A_{i,j,k}^{7\textcolor{myc}{1}}~u_{i-1,j-1,k-1}  &~+~ 
A_{i,j,k}^{8\textcolor{myc}{1}}~u_{i,j-1,k-1}  &~+~ 
A_{i,j,k}^{9\textcolor{myc}{1}}~u_{i+1,j-1,k-1}  &~+  \\
A_{i,j,k}^{10}~u_{i-1,j+1,k\textcolor{myc}{+1}}  &~+~ 
A_{i,j,k}^{11}~u_{i  ,j+1,k\textcolor{myc}{+1}}  &~+~ 
A_{i,j,k}^{12}~u_{i+1,j+1,k\textcolor{myc}{+1}}  &~+  \\  
A_{i,j,k}^{13}~u_{i-1,j\textcolor{myc}{+1},k\textcolor{myc}{+1}}  &~+~ 
A_{i,j,k}^{14}~u_{i  ,j\textcolor{myc}{+1},k\textcolor{myc}{+1}}  &~+~ A_{i,j,k}^{15}~u_{i+1,j\textcolor{myc}{+1},k\textcolor{myc}{+1}}  &~+  \\  
A_{i,j,k}^{16}~u_{i-1,j-1,k\textcolor{myc}{+1}}  &~+~ 
A_{i,j,k}^{17}~u_{i  ,j-1,k\textcolor{myc}{+1}}  &~+~ 
A_{i,j,k}^{18}~u_{i+1,j-1,k\textcolor{myc}{+1}}  &~+  \\
A_{i,j,k}^{19}~u_{i-1,j+1,k+1}  &~+~ 
A_{i,j,k}^{20}~u_{i  ,j+1,k+1}  &~+~ 
A_{i,j,k}^{21}~u_{i+1,j+1,k+1}  &~+  \\  
A_{i,j,k}^{22}~u_{i-1,j\textcolor{myc}{+1},k+1}  &~+~ 
A_{i,j,k}^{23}~u_{i  ,j\textcolor{myc}{+1},k+1}  &~+~ 
A_{i,j,k}^{24}~u_{i+1,j\textcolor{myc}{+1},k+1}  &~+ \\  
A_{i,j,k}^{25}~u_{i-1,j-1,k+1}  &~+~ 
A_{i,j,k}^{26}~u_{i,j-1,k+1}  &~+~ 
A_{i,j,k}^{27}~u_{i+1,j-1,k+1} &=Q_{i,j,k}~, 
\end{alignedat}
\end{equation}
where the RHS is obtained using the following relation
\begin{equation}
\label{eq:q}
\begin{alignedat}{3} 
Q_{i,j,k}&=
q_{i,j,k}^{1\textcolor{myc}{1}}~f_{i-1,j+1,k-1}  &~+~ 
q_{i,j,k}^{2\textcolor{myc}{1}}~f_{i  ,j+1,k-1}  &~+~ 
q_{i,j,k}^{3\textcolor{myc}{1}}~f_{i+1,j+1,k-1}  \\  &~+ 
q_{i,j,k}^{4\textcolor{myc}{1}}~f_{i-1,j\textcolor{myc}{+1},k-1}  &~+~ q_{i,j,k}^{5\textcolor{myc}{1}}~f_{i  ,j\textcolor{myc}{+1},k-1}  &~+~ q_{i,j,k}^{6\textcolor{myc}{1}}~f_{i+1,j\textcolor{myc}{+1},k-1}  \\ &~+  
q_{i,j,k}^{7\textcolor{myc}{1}}~f_{i-1,j-1,k-1}  &~+~ 
q_{i,j,k}^{8\textcolor{myc}{1}}~f_{i,j-1,k-1}  &~+~ 
q_{i,j,k}^{9\textcolor{myc}{1}}~f_{i+1,j-1,k-1}  \\ &~+ 
q_{i,j,k}^{10}~f_{i-1,j+1,k\textcolor{myc}{+1}}  &~+~ 
q_{i,j,k}^{11}~f_{i  ,j+1,k\textcolor{myc}{+1}}  &~+~ 
q_{i,j,k}^{12}~f_{i+1,j+1,k\textcolor{myc}{+1}}  \\ &~+  
q_{i,j,k}^{13}~f_{i-1,j\textcolor{myc}{+1},k\textcolor{myc}{+1}}  &~+~ 
q_{i,j,k}^{14}~f_{i  ,j\textcolor{myc}{+1},k\textcolor{myc}{+1}}  &~+~ q_{i,j,k}^{15}~f_{i+1,j\textcolor{myc}{+1},k\textcolor{myc}{+1}}  \\  &~+ 
q_{i,j,k}^{16}~f_{i-1,j-1,k\textcolor{myc}{+1}}  &~+~ 
q_{i,j,k}^{17}~f_{i  ,j-1,k\textcolor{myc}{+1}}  &~+~ 
q_{i,j,k}^{18}~f_{i+1,j-1,k\textcolor{myc}{+1}}   \\ &~+ 
q_{i,j,k}^{19}~f_{i-1,j+1,k+1}  &~+~ 
q_{i,j,k}^{20}~f_{i  ,j+1,k+1}  &~+~ 
q_{i,j,k}^{21}~f_{i+1,j+1,k+1}  \\ &~+   
q_{i,j,k}^{22}~f_{i-1,j\textcolor{myc}{+1},k+1}  &~+~ 
q_{i,j,k}^{23}~f_{i  ,j\textcolor{myc}{+1},k+1}  &~+~ 
q_{i,j,k}^{24}~f_{i+1,j\textcolor{myc}{+1},k+1}  \\  &~+ 
q_{i,j,k}^{25}~f_{i-1,j-1,k+1}  &~+~ 
q_{i,j,k}^{26}~f_{i,j-1,k+1}  &~+~ 
q_{i,j,k}^{27}~f_{i+1,j-1,k+1}.
\end{alignedat}
\end{equation}
The coefficients for LHS in Eq. (\ref{eq:A27}) and the coefficents for RHS in Eq. (\ref{eq:q}) are given by
\begin{align}
\begin{alignedat}{2}
A_{i,j,k}^{9\alpha+1}&=ar^x_{i}~c^y_{j}~\beta^z_{k}+a^x_{i}~cr^y_{j}~\beta^z_{k}+a^x_{i}~c^y_{j}~\beta r^z_{k}~, ~~~~~~~~~~ &q_{i,j,k}^{9\alpha+1}&=a^x_{i}~c^y_{j}~\beta^z_{k}~,\\
A_{i,j,k}^{9\alpha+2}&=br^x_{i}~c^y_{j}~\beta^z_{k}+b^x_{i}~cr^y_{j}~\beta^z_{k}+b^x_{i}~c^y_{j}~\beta r^z_{k}~, ~~~~~~~~~~ &q_{i,j,k}^{9\alpha+2}&=b^x_{i}~c^y_{j}~\beta^z_{k}~,\\
A_{i,j,k}^{9\alpha+3}&=cr^x_{i}~c^y_{j}~\beta^z_{k}+c^x_{i}~cr^y_{j}~\beta^z_{k}+c^x_{i}~c^y_{j}~\beta r^z_{k}~, ~~~~~~~~~~ &q_{i,j,k}^{9\alpha+3}&=c^x_{i}~c^y_{j}~\beta^z_{k}~,\\
A_{i,j,k}^{9\alpha+4}&=ar^x_{i}~b^y_{j}~\beta^z_{k}+a^x_{i}~br^y_{j}~\beta^z_{k}+a^x_{i}~b^y_{j}~\beta r^z_{k}~, ~~~~~~~~~~ &q_{i,j,k}^{9\alpha+4}&=a^x_{i}~b^y_{j}~\beta^z_{k}~,\\  
A_{i,j,k}^{9\alpha+5}&=br^x_{i}~b^y_{j}~\beta^z_{k}+b^x_{i}~br^y_{j}~\beta^z_{k}+b^x_{i}~b^y_{j}~\beta r^z_{k}~, ~~~~~~~~~~ &q_{i,j,k}^{9\alpha+5}&=b^x_{i}~b^y_{j}~\beta^z_{k}~,\\  
A_{i,j,k}^{9\alpha+6}&=cr^x_{i}~b^y_{j}~\beta^z_{k}+c^x_{i}~br^y_{j}~\beta^z_{k}+c^x_{i}~b^y_{j}~\beta r^z_{k}~, ~~~~~~~~~~ &q_{i,j,k}^{9\alpha+6}&=c^x_{i}~b^y_{j}~\beta^z_{k}~,\\ 
A_{i,j,k}^{9\alpha+7}&=ar^x_{i}~a^y_{j}~\beta^z_{k}+a^x_{i}~ar^y_{j}~\beta^z_{k}+a^x_{i}~a^y_{j}~\beta r^z_{k}~, ~~~~~~~~~~ &q_{i,j,k}^{9\alpha+7}&=a^x_{i}~a^y_{j}~\beta^z_{k}~,\\  
A_{i,j,k}^{9\alpha+8}&=br^x_{i}~a^y_{j}~\beta^z_{k}+b^x_{i}~ar^y_{j}~\beta^z_{k}+b^x_{i}~a^y_{j}~\beta r^z_{k}~, ~~~~~~~~~~ &q_{i,j,k}^{9\alpha+8}&=b^x_{i}~a^y_{j}~\beta^z_{k}~,\\  
A_{i,j,k}^{9\alpha+9}&=cr^x_{i}~a^y_{j}~\beta^z_{k}+c^x_{i}~ar^y_{j}~\beta^z_{k}+c^x_{i}~a^y_{j}~\beta r^z_{k}~. ~~~~~~~~~~ &q_{i,j,k}^{9\alpha+9}&=c^x_{i}~a^y_{j}~\beta^z_{k}~.  
\end{alignedat}
\label{eq:As_qs}
\end{align}
In Eq. (\ref{eq:As_qs}), $\alpha$ is 0, 1 and 2 with corresponding $\beta=a$, $b$ and $c$, respectively. 

\subsection{Treatment of irregular grid points: Fourth-order sharp immersed interface method} \label{sec:cfd_irreg}
In this section, the method to determine the coefficients of the compact scheme stencil at an irregular grid point is presented. Our method falls into the sharp interface category. However, it distinguishes itself from other IIMs 
\begin{figure}[htbp!]
    \centerline{    
    \includegraphics[width=0.98\textwidth]{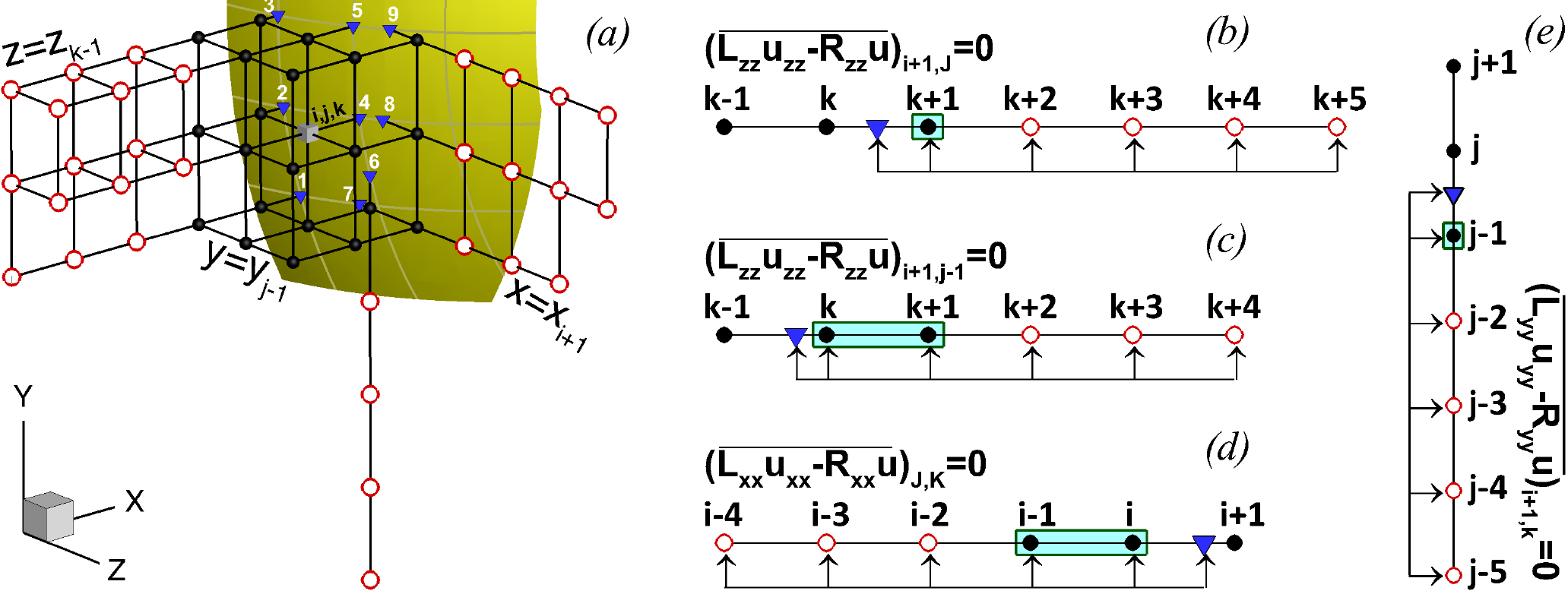}}
    \caption{(\textit{a}) Intersection of 27-point compact FD stencil at irregular grid point ($i,j,k$) with an immersed boundary. Circles are additional grid points used in the modified finite difference operators to maintain the formal fourth-order accuracy. IBI points are numbered from 1 to 9, i.e. $IBI_1,~IBI_2, \cdots,~IBI_9$. Modified finite difference (MFD) operators employed in the $z-$direction along the plane $x=x_{i+1}$ at $y=y_{J}$ with $J=j$ \& $J=j+1$ (\textit{b}) and $y=y_{j-1}$ (\textit{c}). MFD operators used in the $x-$direction corresponding to $IBI_1-IBI_5$ (\textit{d}), and MFD operators employed in the $y$-direction along the plane $x=x_{i+1}$ at $z=z_k$ (\textit{e}). Grid points used on the RHS of the MFD operators are marked with an up-arrow sign. Grid points enclosed by rectangles are the points used in the LHS of the MFD operators.}
    \label{fig:irr_grid_point}
\end{figure}
in that jump corrections are no longer required. The key aspect of the new method is to modify and adjust the compact finite-difference operators, Eqs. (\ref{eq:cfd_lhs_xxyyzz})-(\ref{eq:cfd_rhs_xxyyzz}), when they intersect an immersed boundary to obtain a solution that is sharp across the interface while keeping the fourth-order global accuracy. In particular, for an irregular grid point located at ($i,j,k$), modified FD operators need to be employed along each grid line in the range $x_{i-1}\leq x\leq x_{i+1}$, $y_{j-1}\leq y\leq y_{j+1}$ and $z_{k-1}\leq z\leq z_{k+1}$, if there is an intersection with the immersed boundary. Otherwise, the standard FD operators can be used. For example, in the $y-$direction, in the range $y_{j-1}\leq y\leq y_{j+1}$, check if the grid lines passing through $x=x_{i-1}$, $x=x_i$ and $x=x_{i+1}$ in the planes $z=z_{k-1}$, $z=z_k$ and $z=z_{k+1}$, have any intersection with the immersed boundary. If yes, modified FD operators have to be used along that line, otherwise, the standard FD operators in the $y-$direction given by Eqs. (\ref{eq:cfd_lhs_xxyyzz})-(\ref{eq:cfd_rhs_xxyyzz}) are valid and can therefore be used. Similar approach is employed in the $y-$ and $z-$directions. The modified compact FD stencil will be built based on the combination of the standard and modified FD operators.

For illustration, we consider the $27-$point 3D stencil centered at irregular grid point ($i,j,k$) as shown in Fig. \ref{fig:irr_grid_point}(a). In the $x-$direction, the lines $y=y_{j-1}$, $y=y_j$, and $y=y_{j+1}$ in the plane $z=z_{k-1}$, and the lines $y=y_j$ and $y=y_{j+1}$ in the plane $z=z_k$ do intersect the immersed boundary. In the $y-$direction, the line passing through $x=x_{i+1}$ in the plane $z=z_k$ crosses the immersed boundary. Finally, the lines $y=y_{j-1}$, $y=y_j$, and $y=y_{j+1}$ in the plane $x=x_{i+1}$ do intersect the immersed boundary in the $z-$direction. The FD operators along these lines need to be adjusted to take into account the immersed boundary while maintaining the formal fourth-order accuracy. Eq. (\ref{eq:lhsrhs_xxyyzz}) is rewritten as
\begin{equation}
 \label{eq:lhsrhs_xxyyzz_mod}
\begin{alignedat}{3} 
a^y_j~a^z_k~(\overline{L_{xx}u_{xx}}-\overline{R_{xx}u})_{j-1,k-1} &+ b^y_j~a^z_k~(\overline{L_{xx}u_{xx}}-\overline{R_{xx}u})_{j  ,k-1} &{}+{}& c^y_j~a^z_k~(\overline{L_{xx}u_{xx}}-\overline{R_{xx}u})_{j+1,k-1} &{}+{}& \\
a^y_j~b^z_k~(L_{xx}u_{xx}-R_{xx}u)_{j-1,k\textcolor{myc}{+1}} &+ b^y_j~b^z_k~(\overline{L_{xx}u_{xx}}-\overline{R_{xx}u})_{j  ,k\textcolor{myc}{+1}} &{}+{}& c^y_j~b^z_k~(\overline{L_{xx}u_{xx}}-\overline{R_{xx}u})_{j+1,k\textcolor{myc}{+1}} &{}+{}&  \\
a^y_j~c^z_k~(L_{xx}u_{xx}-R_{xx}u)_{j-1,k+1} &+ b^y_j~c^z_k~(L_{xx}u_{xx}-R_{xx}u)_{j  ,k+1} &{}+{}& c^y_j~c^z_k~(L_{xx}u_{xx}-R_{xx}u)_{j+1,k+1} &{}+{}& \\
a^x_i~a^z_k~(L_{yy}u_{yy}-R_{yy}u)_{i-1,k-1} &+ b^x_i~a^z_k~(L_{yy}u_{yy}-R_{yy}u)_{i  ,k-1} &{}+{}& c^x_i~a^z_k~(\hcancel[red]{L_{yy}u_{yy}-R_{yy}u})_{i+1,k-1} &{}+{}& \\
a^x_i~b^z_k~(L_{yy}u_{yy}-R_{yy}u)_{i-1,k\textcolor{myc}{+1}} &+ b^x_i~b^z_k~(L_{yy}u_{yy}-R_{yy}u)_{i  ,k\textcolor{myc}{+1}} &{}+{}& c^x_i~b^z_k~(\overline{L_{yy}u_{yy}}-\overline{R_{yy}u})_{i+1,k\textcolor{myc}{+1}} &{}+{}& \\
a^x_i~c^z_k~(L_{yy}u_{yy}-R_{yy}u)_{i-1,k+1} &+ b^x_i~c^z_k~(L_{yy}u_{yy}-R_{yy}u)_{i  ,k+1} &{}+{}& c^x_i~c^z_k~(L_{yy}u_{yy}-R_{yy}u)_{i+1,k+1} &{}+{}& \\
a^x_i~a^y_j~(L_{zz}u_{zz}-R_{zz}u)_{i-1,j-1~} &+ b^x_i~a^y_j~(L_{zz}u_{zz}-R_{zz}u)_{i  ,j-1~} &{}+{}& c^x_i~a^y_j~(\overline{L_{zz}u_{zz}}-\overline{R_{zz}u})_{i+1,j-1~} &{}+{}& \\
a^x_i~b^y_j~(L_{zz}u_{zz}-R_{zz}u)_{i-1,j~~~~} &+ b^x_i~b^y_j~(L_{zz}u_{zz}-R_{zz}u)_{i  ,j~~~~} &{}+{}& c^x_i~b^y_j~(\overline{L_{zz}u_{zz}}-\overline{R_{zz}u})_{i+1,j~~~~} &{}+{}& \\
a^x_i~c^y_j~(L_{zz}u_{zz}-R_{zz}u)_{i-1,j+1~} &+ b^x_i~c^y_j~(L_{zz}u_{zz}-R_{zz}u)_{i  ,j+1~} &{}+{}& c^x_i~c^y_j~(\overline{L_{zz}u_{zz}}-\overline{R_{zz}u})_{i+1,j+1~} &{}={}&~0.
\end{alignedat}
\end{equation}

In the above equation, the FD operators with the overbar are modified, while the other operators are the same as those given in Eqs. (\ref{eq:cfd_lhs_xxyyzz})-(\ref{eq:cfd_rhs_xxyyzz}). A key point is that the coefficients used in the modified LHS FD operators ($\overline{L_{xx}}, \overline{L_{yy}}, \overline{L_{zz}}$) in Eq. (\ref{eq:lhsrhs_xxyyzz_mod}) must be kept the same as those used in Eq. (\ref{eq:cfd_lhs_xxyyzz}). However, the coefficients corresponding to the grid points inside the immersed body have to be dropped. Therefore, to maintain the formal fourth order accuracy, additional grid points are needed to determine the coefficients for the modified RHS FD operators ($\overline{R_{xx}}, \overline{R_{yy}}, \overline{R_{zz}}$) in Eq. (\ref{eq:lhsrhs_xxyyzz_mod}). Furthermore, the line passing through $z=z_{k-1}$ in the plane $x=x_{i+1}$ in the $y-$direction falls into the subdomain $\Omega^{-}$, hence, the corresponding FD operators are dropped (struck through in Eq. \ref{eq:lhsrhs_xxyyzz_mod}).

In the $x-$direction along the planes $z=z_{k-1}$ and $z=z_k$ (see Fig. \ref{fig:irr_grid_point}d), $(\overline{L_{xx}u_{xx}})_{J,K}=(\overline{R_{xx}u})_{J,K}$ where
\begin{align}
(\overline{L_{xx}u_{xx}})_{J,K} &= a^x_i~u_{{xx}_{i-1,J,K}} + b^x_i~u_{{xx}_{i,J,K}}~, \label{eq:mod1_top} \\
(\overline{R_{xx}u})_{J,K} &= \overline{ar_{i,\alpha}^{x}}~u_{i-1,J,K} + \overline{br^x_{i,\alpha}}~u_{i,J,K} +\psi_{1,\alpha}~u_{i-2,J,K} +\psi_{2,\alpha}~u_{i-3,J,K} +\psi_{3,\alpha}~u_{i-4,J,K} +\psi_{b,\alpha}~u_{{IBI}_\alpha}~.
\label{eq:mod1}
\end{align}
In Eqs. (\ref{eq:mod1_top})-(\ref{eq:mod1}), $J$ is $j-1$, $j$, and $j+1$ for $K=k-1$ with corresponding $\alpha=1$, $2$ and $3$, respectively. For $K=k$, $J$ is $j$, and $j+1$ with $\alpha=4$ and $5$. Furthermoe, $u_{{IBI}_\alpha}$ is the known function value at the body intercept location $IBI_\alpha$. It can be seen that the term $c^x_i~u_{{xx}_{i+1,J,K}}$ is not included in Eq. (\ref{eq:mod1_top}) since it is inside the immersed body. For the right-hand side operator, Eq. (\ref{eq:mod1}), we use three additional grid points to keep the fourth-order accuracy (see Fig. \ref{fig:irr_grid_point}d). Taylor series expansions about $u_{i,J,K}$ are used to find the coefficients by solving the following system of equations:
\begin{equation}
    \mathcal{M}~\bigg[\overline{br^x_{i,\alpha}}, \overline{ar^x_{i,\alpha}},~\psi_{1,\alpha},~\psi_{2,\alpha},~\psi_{3,\alpha},~\psi_{b,\alpha}\bigg]^{T}=\bigg[0,~0,~2(a^x_i+b^x_i),~6~h_0~a^x_i,~12~h_0^2~a^x_i,~20~h_0^3~a^x_i\bigg]^{T}.
    \label{eq:mat1}
\end{equation}
In the above equation, the coefficient matrix $\mathcal{M}$ is defined as
\begin{equation}
\mathcal{M}=
\begin{bmatrix}
    1   &1      &1      &1      &1      &1     \\[0.5em]
    0   &h_0    &h_1    &h_2    &h_3    &h_b   \\[0.5em]
    0   &h_0^2  &h_1^2  &h_2^2  &h_3^2  &h_b^2 \\[0.5em]
    0   &h_0^3  &h_1^3  &h_2^3  &h_3^3  &h_b^3 \\[0.5em]
    0   &h_0^4  &h_1^4  &h_2^4  &h_3^4  &h_b^4 \\[0.5em] 
    0   &h_0^5  &h_1^5  &h_2^5  &h_3^5  &h_b^5 
\end{bmatrix},
\label{eq:M}
\end{equation}
where $h_0=x_{i-1}-x_i$, $h_1=x_{i-2}-x_i$, $h_2=x_{i-3}-x_i$, $h_3=x_{i-4}-x_i$ and $h_b=x_{{IBI}_\alpha}-x_i$. Note that the coefficients for the RHS in Eq. (\ref{eq:mat1}) are given by Eq. (\ref{eq:app_xxx}). For the stencil passing through the line $x=x_{i+1}$ along $z=z_k$ as shown in Fig. \ref{fig:irr_grid_point}(e), $(\overline{L_{yy}u_{yy}})_{i+1,k}=(\overline{R_{yy}u})_{i+1,k}$ where the modified FD operators are given by
\begin{align}
(\overline{L_{yy}u_{yy}})_{i+1,k} &= a^y_j~u_{{yy}_{i+1,j-1,k}}~, \label{eq:mod2_top} \\
(\overline{R_{yy}u})_{i+1,k} &= \overline{ar_{j}^{y}}~u_{i+1,j-1,k} +\beta_1~u_{i+1,j-2,k} +\beta_2~u_{i+1,j-3,k} +\beta_3~u_{i+1,j-4,k} + \beta_4~u_{i+1,j-5,k} +\beta_b~u_{{IBI}_6}~,
\label{eq:mod2}
\end{align}
$u_{{IBI}_6}$ denoting the known boundary value at IBI location 6, $IBI_6$. One should note that the Taylor expansion about $u_{i+1,j-1,k}$ is used to find the coefficients in Eq. (\ref{eq:mod2}) as follows
\begin{equation}
    \mathcal{M}~\bigg[ \overline{ar^y_{j}},~\beta_1,~\beta_2,~\beta_3,~\beta_4,~\beta_b \bigg]^{T}=\bigg[0,~0,~2~a^y_j,~0,~0,~0 \bigg]^{T},
    \label{eq:mat2}
\end{equation}
where in the matrix $\mathcal{M}$, given by Eq. (\ref{eq:M}), $h_0=y_{j-2}-y_{j-1}$, $h_1=y_{j-3}-y_{j-1}$, $h_2=y_{j-4}-y_{j-1}$, $h_3=y_{j-5}-y_{j-1}$ and $h_b=y_{{IBI}_6}-y_{j-1}$. Finally, for the stencil along the plane $x=x_{i+1}$, the FD scheme takes the form $(\overline{L_{zz}u_{zz}})_{i+1,J}=(\overline{R_{zz}u})_{i+1,J}$
\begin{align}
(\overline{L_{zz}u_{zz}})_{i+1,J} &= b^z_k~u_{{zz}_{i+1,J,k}} + c^z_k~u_{{zz}_{i+1,J,k+1}}~, \label{eq:mod3_top} \\
(\overline{R_{zz}u})_{i+1,J} &= \overline{br_{k}^{z}}~u_{i+1,J,k} + \overline{cr^z_{k,1}}~u_{i+1,J,k+1} +\theta_{1,1}~u_{i+1,J,k+2} +\theta_{2,1}~u_{i+1,J,k+3} +\theta_{3,1}~u_{i+1,J,k+4} +\theta_{b,1}~u_{{IBI}_7}~.
\label{eq:mod3}
\end{align}

\noindent The above equations hold for $J=j-1$ (see Fig. \ref{fig:irr_grid_point}c). For $J=j$ and $J=j+1$, the modified FD operators are given by
\begin{align}
(\overline{L_{zz}u_{zz}})_{i+1,J} &= c^z_k~u_{{zz}_{i+1,j-1,k+1}}~, \label{eq:mod4_top} \\
(\overline{R_{zz}u})_{i+1,J} &= \overline{cr^z_{k,\alpha}}~u_{i+1,J,k+1} +\theta_{1,\alpha}~u_{i+1,J,k+2} +\theta_{2,\alpha}u_{i+1,J,k+3} +\theta_{3,\alpha}u_{i+1,J,k+4} +\theta_{4,\alpha}u_{i+1,J,k+5} +\theta_{b,\alpha}u_{{IBI}_{\alpha+6}},
\label{eq:mod4}
\end{align}
with $\alpha=2$ and $\alpha=3$ for $J=j$ and $J=j+1$, respectively (see Fig. \ref{fig:irr_grid_point}b). Matching the Taylor series coefficients about $u_{i+1,j-1,k}$ in the $z-$direction, the coefficients in Eq. (\ref{eq:mod3}) can be found from
\begin{equation}
\mathcal{M}~\bigg[\overline{br^z_{k}},~\overline{cr^z_{k,1}},~\theta_{1,1},~\theta_{2,1},~\theta_{3,1},~\theta_{b,1}  \bigg]^{T}=
   \bigg[0,~
    0,~2(b^z_k+c^z_k),~6~h_0~c^z_k,~12~h_0^2~c^z_k,~20~h_0^3~c^z_k   \bigg]^{T},
    \label{eq:mat3}
\end{equation}
where $h_0=z_{k+1}-z_k$, $h_1=z_{k+2}-z_k$, $h_2=z_{k+3}-z_k$, $h_3=z_{k+4}-z_k$ and $h_b=z_{{IBI}_7}-z_k$ are used in the matrix $\mathcal{M}$. In Eq. (\ref{eq:mod4}), the coefficients are obtained by matching the Taylor coefficients about $u_{i+1,J,k+1}$
\begin{equation}
\mathcal{M}~\bigg[    \overline{cr^z_{k,\alpha}},~
    \theta_{1,\alpha},~
    \theta_{2,\alpha},~
    \theta_{3,\alpha},~
    \theta_{4,\alpha},~
    \theta_{b,\alpha} \bigg]^{T}=\bigg[ 0,~0,~2~c^z_k,~0,~0,~0\bigg]^{T}, 
    \label{eq:mat4}
\end{equation}
where $h_0=z_{k+2}-z_{k+1}$, $h_1=z_{k+3}-z_{k+1}$, $h_2=z_{k+4}-z_{k+1}$, $h_3=z_{k+5}-z_{k+1}$ and $h_b=z_{{IBI}_\alpha}-z_{k+1}$ are used in the coefficient matrix $\mathcal{M}$ in Eq. (\ref{eq:M}). Applying the above equations to Eq. (\ref{eq:lhsrhs_xxyyzz_mod}), we get the modified compact scheme stencil at the irregular grid point ($i,j,k$)

\begin{equation}
 \label{eq:A27_mod}
  \begin{alignedat}{4}
\overline{A_{i,j,k}^{1\textcolor{myc}{1}}}~u_{i-1,j+1,k-1}  &~+~ 
\overline{A_{i,j,k}^{2\textcolor{myc}{1}}}~u_{i  ,j+1,k-1}  &~+~ 
\hcancel[red]{A_{i,j,k}^{3\textcolor{myc}{1}}~u_{i+1,j+1,k-1}}  &~+  \\  
\overline{A_{i,j,k}^{4\textcolor{myc}{1}}}~u_{i-1,j\textcolor{myc}{+1},k-1}  &~+~ \overline{A_{i,j,k}^{5\textcolor{myc}{1}}}~u_{i  ,j\textcolor{myc}{+1},k-1}  &~+~ \hcancel[red]{A_{i,j,k}^{6\textcolor{myc}{1}}~u_{i+1,j\textcolor{myc}{+1},k-1}}  &~+  \\  
\overline{A_{i,j,k}^{7\textcolor{myc}{1}}}~u_{i-1,j-1,k-1}  &~+~ 
\overline{A_{i,j,k}^{8\textcolor{myc}{1}}}~u_{i,j-1,k-1}  &~+~ 
\hcancel[red]{A_{i,j,k}^{9\textcolor{myc}{1}}~u_{i+1,j-1,k-1}}  &~+  \\
\overline{A_{i,j,k}^{10}}~u_{i-1,j+1,k\textcolor{myc}{+1}}  &~+~ 
\overline{A_{i,j,k}^{11}}~u_{i  ,j+1,k\textcolor{myc}{+1}}  &~+~ 
\hcancel[red]{A_{i,j,k}^{12}~u_{i+1,j+1,k\textcolor{myc}{+1}}}  &~+  \\  
\overline{A_{i,j,k}^{13}}~u_{i-1,j\textcolor{myc}{+1},k\textcolor{myc}{+1}}  &~+~ 
\overline{A_{i,j,k}^{14}}~u_{i  ,j\textcolor{myc}{+1},k\textcolor{myc}{+1}}  &~+~ \hcancel[red]{A_{i,j,k}^{15}~u_{i+1,j\textcolor{myc}{+1},k\textcolor{myc}{+1}}}  &~+  \\  
A_{i,j,k}^{16}~u_{i-1,j-1,k\textcolor{myc}{+1}}  &~+~ 
A_{i,j,k}^{17}~u_{i  ,j-1,k\textcolor{myc}{+1}}  &~+~ 
\overline{A_{i,j,k}^{18}}~u_{i+1,j-1,k\textcolor{myc}{+1}}  &~+  \\
A_{i,j,k}^{19}~u_{i-1,j+1,k+1}  &~+~ 
A_{i,j,k}^{20}~u_{i  ,j+1,k+1}  &~+~ 
\overline{A_{i,j,k}^{21}}~u_{i+1,j+1,k+1}  &~+  \\  
A_{i,j,k}^{22}~u_{i-1,j\textcolor{myc}{+1},k+1}  &~+~ 
A_{i,j,k}^{23}~u_{i  ,j\textcolor{myc}{+1},k+1}  &~+~ 
\overline{A_{i,j,k}^{24}}~u_{i+1,j\textcolor{myc}{+1},k+1}  &~+ \\  
A_{i,j,k}^{25}~u_{i-1,j-1,k+1}  &~+~ 
A_{i,j,k}^{26}~u_{i,j-1,k+1}  &~+~ 
\overline{A_{i,j,k}^{27}}~u_{i+1,j-1,k+1} &=\overline{Q_{i,j,k}}-\mathcal{B}_{i,j,k}-\mathcal{C}_{i,j,k}~, 
\end{alignedat}
\end{equation}
with the modified coefficients  
\begin{align}
\begin{alignedat}{2}
\overline{A_{i,j,k}^{1}}&=\overline{ar^x_{i,3}}~c^y_{j}~a^z_{k}+a^x_{i}~cr^y_{j}~a^z_{k}+a^x_{i}~c^y_{j}~ar^z_{k}~, ~~~~~~~~~~ &\overline{A_{i,j,k}^{11}}&=\overline{br^x_{i,5}}~c^y_{j}~b^z_{k}+b^x_{i}~cr^y_{j}~b^z_{k}+b^x_{i}~c^y_{j}~br^z_{k}~,\\
\overline{A_{i,j,k}^{2}}&=\overline{br^x_{i,3}}~c^y_{j}~a^z_{k}+b^x_{i}~cr^y_{j}~a^z_{k}+b^x_{i}~c^y_{j}~ar^z_{k}~, ~~~~~~~~~~ &\overline{A_{i,j,k}^{13}}&=\overline{ar^x_{i,4}}~b^y_{j}~b^z_{k}+a^x_{i}~br^y_{j}~b^z_{k}+a^x_{i}~b^y_{j}~br^z_{k}~,\\
\overline{A_{i,j,k}^{4}}&=\overline{ar^x_{i,2}}~b^y_{j}~a^z_{k}+a^x_{i}~br^y_{j}~a^z_{k}+a^x_{i}~b^y_{j}~ar^z_{k}~, ~~~~~~~~~~ &\overline{A_{i,j,k}^{14}}&=\overline{br^x_{i,4}}~b^y_{j}~b^z_{k}+b^x_{i}~br^y_{j}~b^z_{k}+b^x_{i}~b^y_{j}~br^z_{k}~,\\
\overline{A_{i,j,k}^{5}}&=\overline{br^x_{i,2}}~b^y_{j}~a^z_{k}+b^x_{i}~br^y_{j}~a^z_{k}+b^x_{i}~b^y_{j}~ar^z_{k}~, ~~~~~~~~~~ &\overline{A_{i,j,k}^{18}}&=cr^x_{i}~a^y_{j}~b^z_{k}~+c^x_{i}~\overline{ar^y_{j}}~b^z_{k}+c^x_{i}~a^y_{j}~\overline{br^z_{k}}~,\\
\overline{A_{i,j,k}^{7}}&=\overline{ar^x_{i,1}}~a^y_{j}~a^z_{k}+a^x_{i}~ar^y_{j}~a^z_{k}+a^x_{i}~a^y_{j}~ar^z_{k}~, ~~~~~~~~~~ &\overline{A_{i,j,k}^{21}}&=cr^x_{i}~c^y_{j}~c^z_{k}~+c^x_{i}~cr^y_{j}~c^z_{k}~+c^x_{i}~c^y_{j}~\overline{cr^z_{k,3}}~,\\
\overline{A_{i,j,k}^{8}}&=\overline{br^x_{i,1}}~a^y_{j}~a^z_{k}+b^x_{i}~ar^y_{j}~a^z_{k}+b^x_{i}~a^y_{j}~ar^z_{k}~, ~~~~~~~~~~ &\overline{A_{i,j,k}^{24}}&=cr^x_{i}~b^y_{j}~c^z_{k}~+c^x_{i}~br^y_{j}~c^z_{k}~+c^x_{i}~b^y_{j}~\overline{cr^z_{k,2}}~,\\
\overline{A_{i,j,k}^{10}}&=\overline{ar^x_{i,5}}~c^y_{j}~b^z_{k}+a^x_{i}~cr^y_{j}~b^z_{k}+a^x_{i}~c^y_{j}~br^z_{k}~, ~~~~~~~~~~ &\overline{A_{i,j,k}^{27}}&=cr^x_{i}~a^y_{j}~c^z_{k}~+c^x_{i}~ar^y_{j}~c^z_{k}~+c^x_{i}~a^y_{j}~\overline{cr^z_{k,1}}~,
\end{alignedat}
\label{eeq:As_mod}
\end{align}

The modified RHS term $\overline{Q_{i,j,k}}$ is the same as the one in Eq. (\ref{eq:q}) except that all the coefficients corresponding to the grid points that lie inside the immersed body are set to zero. In Eq. (\ref{eq:A27_mod}), $\mathcal{B}_{i,j,k}$ includes the known function values at the immersed boundary intercept points,
\begin{equation}
    \begin{alignedat}{3}
    \mathcal{B}_{i,j,k}=a^y_j~a^z_k~\psi_{b,1}~u_{{IBI}_1} &{}+{} b^y_j~a^z_k~\psi_{b,2}~u_{{IBI}_2} &{}+{} c^y_j~a^z_k~\psi_{b,3}~u_{{IBI}_3} &{}+{} b^y_j~b^z_k~\psi_{b,4}~u_{{IBI}_4} + \\
c^y_j~b^z_k~\psi_{b,5}~u_{{IBI}_5} &{}+{} c^x_i~b^z_k~~~\beta_{b}~~u_{{IBI}_6} &{}+{} c^x_i~a^y_j~\theta_{b,1}~u_{{IBI}_7} &{}+{} c^x_i~b^y_j~\theta_{b,2}~u_{{IBI}_8} + c^x_i~c^y_j~\theta_{b,3}~u_{{IBI}_9}~,   
    \end{alignedat}
    \label{eq:bc}
\end{equation}
and $\mathcal{C}_{i,j,k}$ contains the additional points used to keep the fourth-order formal accuracy
\begin{align}
    \begin{alignedat}{3}
    \mathcal{C}_{i,j,k}&=
       a^y_j~a^z_k\big(\psi_{1,1}~u_{i-2,j-1,k-1}                  &+ \psi_{2,1}~u_{i-3,j-1,k-1} &+ \psi_{3,1}~u_{i-4,j-1,k-1}  \big) \\
    &~+b^y_j~a^z_k\big(\psi_{1,2}~u_{i-2,j\textcolor{myc}{+1},k-1} &+ \psi_{2,2}~u_{i-3,j\textcolor{myc}{+1},k-1} &+ \psi_{3,2}~u_{i-4,j\textcolor{myc}{+1},k-1}  \big) \\
    &~+c^y_j~a^z_k\big(\psi_{1,3}~u_{i-2,j+1,k-1}                  &+ \psi_{2,3}~u_{i-3,j+1,k-1} &+ \psi_{3,3}~u_{i-4,j+1,k-1}  \big) \\
    &~+b^y_j~b^z_k\big(\psi_{1,4}~u_{i-2,j\textcolor{myc}{+1},k\textcolor{myc}{+1}} &+ \psi_{2,4}~u_{i-3,j\textcolor{myc}{+1},k\textcolor{myc}{+1}} &+ \psi_{3,4}~u_{i-4,j\textcolor{myc}{+1},k\textcolor{myc}{+1}}  \big) \\
    &~+c^y_j~b^z_k\big(\psi_{1,5}~u_{i-2,j+1,k\textcolor{myc}{+1}} &+ \psi_{2,5}~u_{i-3,j+1,k\textcolor{myc}{+1}} &+ \psi_{3,5}~u_{i-4,j+1,k\textcolor{myc}{+1}}  \big) \\
    &~+c^x_i~a^y_j\big(\theta_{1,1}~u_{i+1,j-1,k+2} &+ \theta_{2,1}~u_{i+1,j-1,k+3} &+ \theta_{3,1}~u_{i+1,j-1,k+4} \big) \\
    &~+c^x_i~b^y_j\big(\theta_{1,2}~u_{i+1,j\textcolor{myc}{+1},k+2} &+ \theta_{2,2}~u_{i+1,j\textcolor{myc}{+1},k+3} &+ \theta_{3,2}~u_{i+1,j\textcolor{myc}{+1},k+4} + \theta_{4,2}~u_{i+1,j\textcolor{myc}{+1},k+5} \big) \\
    &~+c^x_i~c^y_j\big(\theta_{1,3}~u_{i+1,j+1,k+2} &+ \theta_{2,3}~u_{i+1,j+1,k+3} &+ \theta_{3,3}~u_{i+1,j+1,k+4} + \theta_{4,3}~u_{i+1,j+1,k+5} \big) \\
    &~+c^x_i~b^z_k\big(\beta_{1}~~~u_{i+1,j-2,k\textcolor{myc}{+1}} &+~\beta_{2}~~u_{i+1,j-3,k\textcolor{myc}{+1}} &+ \beta_{3}~~~u_{i+1,j-4,k\textcolor{myc}{+1}} + ~\beta_{4}~~u_{i+1,j-5,k\textcolor{myc}{+1}}  \big)~.
    \end{alignedat}
    \label{eq:cor}
\end{align}
 
It is worth noting that the coefficients without the overbar in Eq. (\ref{eq:A27_mod}) are the same as the coefficients obtained for the regular grid points in Eq. (\ref{eq:As_qs}). It should be noted that the coefficients for the solid grid points are not used in Eq. (\ref{eq:A27_mod}) since the function value is zero at those grid points (struck through in Eq. \ref{eq:A27_mod}).

\subsection{Special case: multiple intersections with immersed boundary} \label{sec:cfd_irr_spec}
\begin{figure}[htbp!]
    \centerline{    
    \includegraphics[width=0.9\textwidth]{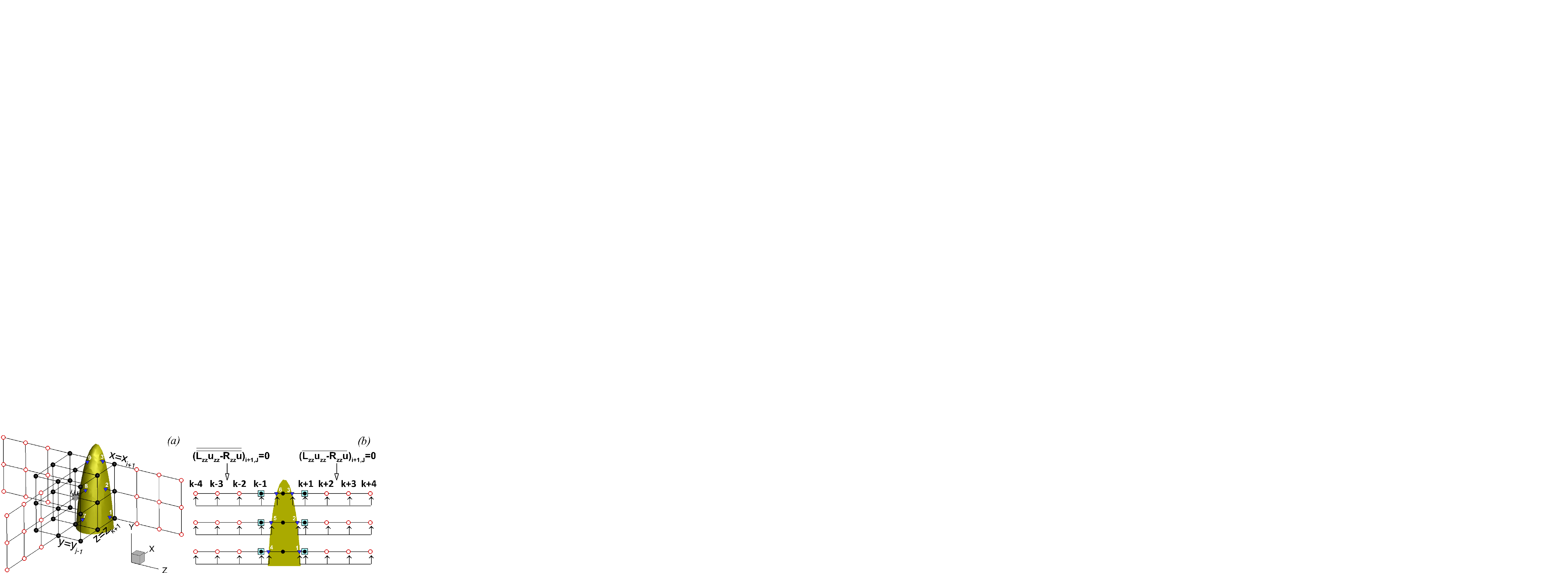}}
    \caption{(\textit{a})
    Schematic of double intersections with immersed boundary for irregular grid point centered at ($i,j,k$). Points of intersection with the immersed boundary are numbered from 1 to 9. (\textit{b}) Modified finite difference operators employed in the $z-$direction along the plane $x=x_{i+1}$ at $y=y_{J}$ with $J=j-1$, $J=j$ and $J=j+1$. For the descriptions regarding the colored rectangles and marked points, see caption of Fig. \ref{fig:irr_grid_point}.}
    \label{fig:irr_grid_case2}
\end{figure}

In the previous section, the immersed boundary has only one intersection with each grid line in the range of $x_{i-1}\leq x\leq x_{i+1}$, $y_{j-1}\leq y\leq y_{j+1}$ and $z_{k-1}\leq z\leq z_{k+1}$. In many practical applications, however, the immersed boundary $\Gamma$ can cross the grid lines more than once as shown in Fig. \ref{fig:irr_grid_case2}, for example. In this case, the compact finite-difference stencil centered at the irregular grid point ($i,j,k$) takes the following form:
\begin{equation}
 \label{eq:lhsrhs_xxyyzz_mod_case2}
\begin{alignedat}{3} 
a^y_j~a^z_k~(L_{xx}u_{xx}-R_{xx}u)_{j-1,k-1} &+ b^y_j~a^z_k~(L_{xx}u_{xx}-R_{xx}u)_{j  ,k-1} &{}+{}& c^y_j~a^z_k~(L_{xx}u_{xx}-R_{xx}u)_{j+1,k-1} &{}+{}& \\
a^y_j~b^z_k~(\overline{L_{xx}u_{xx}}-\overline{R_{xx}u})_{j-1,k\textcolor{myc}{+1}} &+ b^y_j~b^z_k~(\overline{L_{xx}u_{xx}}-\overline{R_{xx}u})_{j  ,k\textcolor{myc}{+1}} &{}+{}& c^y_j~b^z_k~(\overline{L_{xx}u_{xx}}-\overline{R_{xx}u})_{j+1,k\textcolor{myc}{+1}} &{}+{}&  \\
a^y_j~c^z_k~(L_{xx}u_{xx}-R_{xx}u)_{j-1,k+1} &+ b^y_j~c^z_k~(L_{xx}u_{xx}-R_{xx}u)_{j  ,k+1} &{}+{}& c^y_j~c^z_k~(L_{xx}u_{xx}-R_{xx}u)_{j+1,k+1} &{}+{}& \\
a^x_i~a^z_k~(L_{yy}u_{yy}-R_{yy}u)_{i-1,k-1} &+ b^x_i~a^z_k~(L_{yy}u_{yy}-R_{yy}u)_{i  ,k-1} &{}+{}& c^x_i~a^z_k~(L_{yy}u_{yy}-R_{yy}u)_{i+1,k-1} &{}+{}& \\
a^x_i~b^z_k~(L_{yy}u_{yy}-R_{yy}u)_{i-1,k\textcolor{myc}{+1}} &+ b^x_i~b^z_k~(L_{yy}u_{yy}-R_{yy}u)_{i  ,k\textcolor{myc}{+1}} &{}+{}& c^x_i~b^z_k~(\hcancel[red]{L_{yy}u_{yy}-R_{yy}u})_{i+1,k\textcolor{myc}{+1}} &{}+{}& \\
a^x_i~c^z_k~(L_{yy}u_{yy}-R_{yy}u)_{i-1,k+1} &+ b^x_i~c^z_k~(L_{yy}u_{yy}-R_{yy}u)_{i  ,k+1} &{}+{}& c^x_i~c^z_k~(L_{yy}u_{yy}-R_{yy}u)_{i+1,k+1} &{}+{}& \\
a^x_i~a^y_j~(L_{zz}u_{zz}-R_{zz}u)_{i-1,j-1~} &+ b^x_i~a^y_j~(L_{zz}u_{zz}-R_{zz}u)_{i  ,j-1~} &{}+{}& c^x_i~a^y_j~(\widehat{L_{zz}u_{zz}}-\widehat{R_{zz}u})_{i+1,j-1~} &{}+{}& \\
a^x_i~b^y_j~(L_{zz}u_{zz}-R_{zz}u)_{i-1,j~~~~} &+ b^x_i~b^y_j~(L_{zz}u_{zz}-R_{zz}u)_{i  ,j~~~~} &{}+{}& c^x_i~b^y_j~(\widehat{L_{zz}u_{zz}}-\widehat{R_{zz}u})_{i+1,j~~~~} &{}+{}& \\
a^x_i~c^y_j~(L_{zz}u_{zz}-R_{zz}u)_{i-1,j+1~} &+ b^x_i~c^y_j~(L_{zz}u_{zz}-R_{zz}u)_{i  ,j+1~} &{}+{}& c^x_i~c^y_j~(\widehat{L_{zz}u_{zz}}-\widehat{R_{zz}u})_{i+1,j+1~} &{}={}&~0.
\end{alignedat}
\end{equation}

According to Fig. \ref{fig:irr_grid_case2}(a), the line passing though $z=z_k$ in the plane $x=x_{i+1}$ in the $y-$direction falls into the subdomain $\Omega^{-}$, hence, the corresponding FD operators are dropped (struck through in Eq. \ref{eq:lhsrhs_xxyyzz_mod_case2}). In the $x-$direction, the plane passing through $z=z_k$ crosses the immersed boundary at the $IBI_7$, $IBI_8$ and $IBI_9$. Along this plane, $(\overline{L_{xx}u_{xx}}-\overline{R_{xx}u})_{J,k}=0$ where

\begin{align}
(\overline{L_{xx}u_{xx}})_{J,k} &= a^x_i~u_{{xx}_{i-1,J,k}} + b^x_i~u_{{xx}_{i,J,K}}~, \label{eq:mod1_top_case2} \\
(\overline{R_{xx}u})_{J,k} &= \overline{ar_{i,\alpha}^{x}}~u_{i-1,J,k} + \overline{br^x_{i,\alpha}}~u_{i,J,k} +\psi_{1,\alpha}~u_{i-2,J,k} +\psi_{2,\alpha}~u_{i-3,J,k} +\psi_{3,\alpha}~u_{i-4,J,k} +\psi_{b,\alpha}~u_{{IBI}_{\alpha+6}}~.
\label{eq:mod1_case2}
\end{align}
In Eqs. (\ref{eq:mod1_top_case2})-(\ref{eq:mod1_case2}), $J$ is $j-1$, $j$ and $j+1$ with corresponding $\alpha=1$, $2$ and $3$, respectively. The coefficients in Eq. (\ref{eq:mod1_case2}) can be found from matching the Taylor series coefficients about $u_{i,J,k}$ in the $x-$direction (see Eq. \ref{eq:mat1}). In the $z-$direction, however, the lines $y=y_{j-1}$, $y=y_j$ and $y=y_{j+1}$ in the plane $x=x_{i+1}$ intersect the immersed boundary at more than one location as shown in Fig. \ref{fig:irr_grid_case2}(b), namely $IBI_1$ \& $IBI_4$ along $y=y_{j-1}$, $IBI_2$ \& $IBI_5$ along $y=y_{j}$ and $IBI_3$ \& $IBI_6$ along $y=y_{j+1}$. Therefore, special care is needed here to modify FD operators along these lines. 

For the stencil passing through the plane $x=x_{i+1}$ in the $z-$direction, $(\widehat{L_{zz}u_{zz}})_{i+1,J}=(\widehat{R_{zz}u})_{i+1,J}$ where $(\widehat{L_{zz}u_{zz}})_{i+1,J}=(\overline{L_{zz}u_{zz}})_{i+1,J}+(\overline{\overline{L_{zz}u_{zz}}})_{i+1,J}$ and $(\widehat{R_{zz}u})_{i+1,J}=(\overline{R_{zz}u})_{i+1,J}+(\overline{\overline{R_{zz}u}})_{i+1,J}$, with $J=j-1$, $J=j$ and $J=j+1$. The modified FD operators are given by
\begin{align}
(\overline{L_{zz}u_{zz}})_{i+1,J} &= c^z_k~u_{{zz}_{i+1,J,k+1}}~, \label{eq:mod2_top_case2} \\
(\overline{R_{zz}u})_{i+1,J} &= \overline{cr_{k,\alpha}^{z}}~u_{i+1,J,k+1} +\beta_{1,\alpha}~u_{i+1,J,k+2} +\beta_{2,\alpha}~u_{i+1,J,k+3} +\beta_{3,\alpha}~u_{i+1,J,k+4} +\beta_{b,\alpha}~u_{{IBI}_\alpha}~,
\label{eq:mod2_case2}
\end{align}
and
\begin{align}
(\overline{\overline{L_{zz}u_{zz}}})_{i+1,J} &= a^z_k~u_{{zz}_{i+1,J,k-1}}~, \label{eq:mod2_top_case22} \\
(\overline{\overline{R_{zz}u}})_{i+1,J} &= \overline{\overline{ar_{k,\alpha}^{z}}}~u_{i+1,J,k-1} +\theta_{1,\alpha}~u_{i+1,J,k-2} +\theta_{2,\alpha}~u_{i+1,J,k-3} +\theta_{3,\alpha}~u_{i+1,J,k-4} +\theta_{b,\alpha}~u_{{IBI}_{\alpha+3}}~,
\label{eq:mod2_case22}
\end{align}
with $\alpha=1$, $2$ and $3$ for $J=j-1$, $j$ and $j+1$, respectively. Here, the Taylor series expansions in the $z-$direction about $u_{i+1,J,k+1}$ and $u_{i+1,J,k-1}$ are used to find the coefficients in Eqs. (\ref{eq:mod2_case2}) and (\ref{eq:mod2_case22}), respectively. Using the original FD operators provided by Eqs. (\ref{eq:cfd_lhs_xxyyzz})-(\ref{eq:cfd_rhs_xxyyzz}) for the unmodified operators in Eq. (\ref{eq:lhsrhs_xxyyzz_mod_case2}) and applying the modified FD operators given by Eqs. (\ref{eq:mod1_top_case2})-(\ref{eq:mod2_case22}) to Eq. (\ref{eq:lhsrhs_xxyyzz_mod_case2}), the modified compact scheme stencil at the irregular grid point ($i,j,k$) can be obtained, 
\begin{equation}
 \label{eq:A27_mod_case2}
  \begin{alignedat}{4}
{A_{i,j,k}^{1\textcolor{myc}{1}}}~u_{i-1,j+1,k-1}  &~+~ 
{A_{i,j,k}^{2\textcolor{myc}{1}}}~u_{i  ,j+1,k-1}  &~+~ 
\overline{A_{i,j,k}^{3\textcolor{myc}{1}}}~u_{i+1,j+1,k-1}  &~+  \\  
{A_{i,j,k}^{4\textcolor{myc}{1}}}~u_{i-1,j\textcolor{myc}{+1},k-1}  &~+~ {A_{i,j,k}^{5\textcolor{myc}{1}}}~u_{i  ,j\textcolor{myc}{+1},k-1}  &~+~ \overline{A_{i,j,k}^{6\textcolor{myc}{1}}}~u_{i+1,j\textcolor{myc}{+1},k-1}  &~+  \\  
{A_{i,j,k}^{7\textcolor{myc}{1}}}~u_{i-1,j-1,k-1}  &~+~ 
{A_{i,j,k}^{8\textcolor{myc}{1}}}~u_{i,j-1,k-1}  &~+~ 
\overline{A_{i,j,k}^{9\textcolor{myc}{1}}}~u_{i+1,j-1,k-1}  &~+  \\
\overline{A_{i,j,k}^{10}}~u_{i-1,j+1,k\textcolor{myc}{+1}}  &~+~ 
\overline{A_{i,j,k}^{11}}~u_{i  ,j+1,k\textcolor{myc}{+1}}  &~+~ 
\hcancel[red]{A_{i,j,k}^{12}~u_{i+1,j+1,k\textcolor{myc}{+1}}}  &~+  \\  
\overline{A_{i,j,k}^{13}}~u_{i-1,j\textcolor{myc}{+1},k\textcolor{myc}{+1}}  &~+~ 
\overline{A_{i,j,k}^{14}}~u_{i  ,j\textcolor{myc}{+1},k\textcolor{myc}{+1}}  &~+~ \hcancel[red]{A_{i,j,k}^{15}~u_{i+1,j\textcolor{myc}{+1},k\textcolor{myc}{+1}}}  &~+  \\  
\overline{A_{i,j,k}^{16}}~u_{i-1,j-1,k\textcolor{myc}{+1}}  &~+~ 
\overline{A_{i,j,k}^{17}}~u_{i  ,j-1,k\textcolor{myc}{+1}}  &~+~ 
\hcancel[red]{A_{i,j,k}^{18}~u_{i+1,j-1,k\textcolor{myc}{+1}}}  &~+  \\
A_{i,j,k}^{19}~u_{i-1,j+1,k+1}  &~+~ 
A_{i,j,k}^{20}~u_{i  ,j+1,k+1}  &~+~ 
\overline{A_{i,j,k}^{21}}~u_{i+1,j+1,k+1}  &~+  \\  
A_{i,j,k}^{22}~u_{i-1,j\textcolor{myc}{+1},k+1}  &~+~ 
A_{i,j,k}^{23}~u_{i  ,j\textcolor{myc}{+1},k+1}  &~+~ 
\overline{A_{i,j,k}^{24}}~u_{i+1,j\textcolor{myc}{+1},k+1}  &~+ \\  
A_{i,j,k}^{25}~u_{i-1,j-1,k+1}  &~+~ 
A_{i,j,k}^{26}~u_{i,j-1,k+1}  &~+~ 
\overline{A_{i,j,k}^{27}}~u_{i+1,j-1,k+1} &=\overline{Q_{i,j,k}}-\mathcal{B}_{i,j,k}-\mathcal{C}_{i,j,k}~, 
\end{alignedat}
\end{equation}
The modified coefficients are
\begin{align}
\begin{alignedat}{2}
\overline{A_{i,j,k}^{3}}&=cr^x_{i}~c^y_{j}~a^z_{k}+c^x_{i}~cr^y_{j}~a^z_{k}+c^x_{i}~c^y_{j}~\overline{\overline{ar^z_{k,3}}}~, ~~~~~~~~~~ &
\overline{A_{i,j,k}^{14}}&=\overline{br^x_{i,2}}~b^y_{j}~b^z_{k}+b^x_{i}~br^y_{j}~b^z_{k}+b^x_{i}~b^y_{j}~br^z_{k}~,\\
\overline{A_{i,j,k}^{6}}&=cr^x_{i}~b^y_{j}~a^z_{k}+c^x_{i}~br^y_{j}~a^z_{k}+c^x_{i}~b^y_{j}~\overline{\overline{ar^z_{k,2}}}~, ~~~~~~~~~~ &
\overline{A_{i,j,k}^{16}}&=\overline{ar^x_{i,1}}~a^y_{j}~b^z_{k}+a^x_{i}~ar^y_{j}~b^z_{k}+a^x_{i}~a^y_{j}~br^z_{k}~,\\
\overline{A_{i,j,k}^{9}}&=cr^x_{i}~a^y_{j}~a^z_{k}+c^x_{i}~ar^y_{j}~a^z_{k}+c^x_{i}~a^y_{j}~\overline{\overline{ar^z_{k,1}}}~, ~~~~~~~~~~ &
\overline{A_{i,j,k}^{17}}&=\overline{br^x_{i,1}}~a^y_{j}~b^z_{k}+b^x_{i}~ar^y_{j}~b^z_{k}+b^x_{i}~a^y_{j}~br^z_{k}~,\\
\overline{A_{i,j,k}^{10}}&=\overline{ar^x_{i,3}}~c^y_{j}~b^z_{k}+a^x_{i}~cr^y_{j}~b^z_{k}+a^x_{i}~c^y_{j}~br^z_{k}~, ~~~~~~~~~~ &
\overline{A_{i,j,k}^{21}}&=cr^x_{i}~c^y_{j}~c^z_{k}~+c^x_{i}~cr^y_{j}~c^z_{k}~+c^x_{i}~c^y_{j}~\overline{cr^z_{k,3}}~,\\
\overline{A_{i,j,k}^{11}}&=\overline{br^x_{i,3}}~c^y_{j}~b^z_{k}+b^x_{i}~cr^y_{j}~b^z_{k}+b^x_{i}~c^y_{j}~br^z_{k}~, ~~~~~~~~~~ &\overline{A_{i,j,k}^{24}}&=cr^x_{i}~b^y_{j}~c^z_{k}+c^x_{i}~br^y_{j}~c^z_{k}+c^x_{i}~b^y_{j}~\overline{cr^z_{k,2}}~,\\
\overline{A_{i,j,k}^{13}}&=\overline{ar^x_{i,2}}~b^y_{j}~b^z_{k}+a^x_{i}~br^y_{j}~b^z_{k}+a^x_{i}~b^y_{j}~br^z_{k}~, ~~~~~~~~~~ &
\overline{A_{i,j,k}^{27}}&=cr^x_{i}~a^y_{j}~c^z_{k}~+c^x_{i}~ar^y_{j}~c^z_{k}~+c^x_{i}~a^y_{j}~\overline{cr^z_{k,1}}~.
\end{alignedat}
\label{eeq:As_mod_case2}
\end{align}
The correction term, $\mathcal{C}_{i,j,k}$, is given by
\begin{equation}
    \begin{alignedat}{3}
    \mathcal{C}_{i,j,k}&=c^x_i~a^y_j~ \sum_{n=1}^{3} \beta_{n,1}~u_{i+1,j-1,k+1+n} &~+c^x_i~b^y_j~ \sum_{n=1}^{3} \beta_{n,2}~u_{i+1,j  ,k+1+n} &~+c^x_i~c^y_j~ \sum_{n=1}^{3} \beta_{n,3}~u_{i+1,j+1,k+1+n} \\
 &~+c^x_i~a^y_j~\sum_{n=1}^{3} \theta_{n,1}~u_{i+1,j-1,k-1-n} &~+c^x_i~b^y_j~\sum_{n=1}^{3} \theta_{n,2}~u_{i+1,j  ,k-1-n} &~+c^x_i~c^y_j~\sum_{n=1}^{3} \theta_{n,3}~u_{i+1,j+1,k-1-n} \\
     &~+a^y_j~b^z_k~ \sum_{n=1}^{3} \psi_{n,1}~u_{i-1-n,j-1,k} &~+b^y_j~b^z_k~ \sum_{n=1}^{3} \psi_{n,2}~u_{i-1-n,j,k\textcolor{myc}{+1}} &~+c^y_j~b^z_k~ \sum_{n=1}^{3} \psi_{n,3}~u_{i-1-n,j+1,k}~.
    \end{alignedat}
    \label{eq:cor_case2}
\end{equation}
The known function values at the points of intersection with the immersed boundary are included in $\mathcal{B}_{i,j,k}$,
\begin{equation}
    \begin{alignedat}{3}
    \mathcal{B}_{i,j,k}&=c^x_i~a^y_j~\beta_{b,1}~u_{{IBI}_1} &~+c^x_i~b^y_j~\beta_{b,2}~u_{{IBI}_2}  &~+c^x_i~c^y_j~\beta_{b,3}~u_{{IBI}_3}  \\
 &~+c^x_i~a^y_j~\theta_{b,1}~u_{{IBI}_4}  &~+c^x_i~b^y_j~\theta_{b,2}~u_{{IBI}_5}  &~+c^x_i~c^y_j~\theta_{b,3}~u_{{IBI}_6} \\
 &~+a^y_j~b^z_k~\psi_{b,1}~u_{{IBI}_7} &~+b^y_j~b^z_k~\psi_{b,2}~u_{{IBI}_8} &~+c^y_j~b^z_k~\psi_{b,3}~u_{{IBI}_9}~.
    \end{alignedat}
    \label{eq:qb_case2}
\end{equation}

\subsection{Order of original and modified compact finite-difference stencils} \label{sec:order}
In this section, the order of accuracy of the compact FD stencils for regular and irregular grid points developed in Sections \ref{sec:cfd_reg}$-$\ref{sec:cfd_irr_spec}, is formally verified. It is worth noting that all the FD operators employed to construct the original and modified compact FD stencils, are based on the desired fourth-order compact finite-difference approximations. Using the fact that the eigenfunctions of the exact Laplace operator are known to be $\phi_{k,l,m}=cos(kx)~cos(ly)~cos(mz)$ with corresponding eigenvalues $\lambda=-(k^2+l^2+m^2)$, one can estimate the order of discrete FD operators where eigenvalues change according to the order of the accuracy of finite-difference approximations \citep{steffen,sutmann}. In the derivation of the compact FD stencils, one operator acts on the field $u$, and one operator operates on the source term $f$, i.e. $A_h~\boldsymbol{u}=Q_h~\boldsymbol{f}$, where $A_h$ and $Q_h$ are discrete operators, see Eqs. (\ref{eq:A27})-(\ref{eq:q}) for example. Therefore, a generalized eigenvalue problem, $A_h~\boldsymbol{u}=\lambda~Q_h~\boldsymbol{u}$, has to be solved. As pointed out by \citet{steffen}, both $A_h$ and $Q_h$ have the desired property that eigenfunctions are sampled continuous eigenvectors. As a result, only the eigenvalues need to be considered. For practical purposes one may insert the eigenfunctions of the exact operator and estimate the order of FD approximations in $h=max(dx,dy,dz)$, i.e.,
\begin{equation}
   \big[A_h~\phi_{k,l,m}+(k^2+l^2+m^2)~Q_h~\phi_{k,l,m}\big]~\phi^{-1}_{k,l,m}=O(h^n)~,
    \label{geign_order}
\end{equation}
where $n$ is the order of the difference approximation. Without loss of generality, it is assumed that the grid is uniform in each direction, however, the grid spacing in each direction could be different, i.e. $dx\neq dy\neq dz$. Furthermore, the locations of the points of intersection with the immersed boundary are estimated according to Figs. (\ref{fig:irr_grid_point})-(\ref{fig:irr_grid_case2}). Applying the FD coefficients developed in Sections \ref{sec:cfd_reg}$-$\ref{sec:cfd_irr_spec}, provides the order of the compact FD stencils for a regular grid point, Eq. (\ref{eq:A27}), and irregular grid points, Eqs. (\ref{eq:A27_mod}) and (\ref{eq:A27_mod_case2}), as follows:
\begin{equation}
\begin{aligned}
\varepsilon&=\frac{1}{12000}\bigg(144~k^6 dx^4+144~l^6 dy^4+144~m^6 dz^4 \bigg) +H.O.T~, \\
\varepsilon&=\frac{1}{12000}\bigg(~~83~k^6 dx^4+138~l^6 dy^4+138~m^6 dz^4 \bigg) +H.O.T~, \\
\varepsilon&=\frac{1}{12000}\bigg(840~k^6 dx^4+134~l^6 dy^4+144~m^6 dz^4 \bigg) +H.O.T~,
\end{aligned}
    \label{eq:order_FD}
\end{equation}
where H.O.T refers to higher order terms. For all the FD stencils developed for the regular and irregular grid points, the expected fourth-order accuracy is recovered.

\section{Solution strategy} \label{sec:sol}
The discretization of the 3D Poisson equation on a uniform/nonuniform grid with $nx\times ny\times nz$ points  leads to a set of $(nx-2)\times (ny-2)\times (nz-2)$ linear equations for all unknown function $u_{i,j,k}$ inside the computational domain, $2\leq i \leq nx-1$, $2\leq j\leq ny-1$ and $2\leq k\leq nz-1$. The system of linear equations can be represented in the matrix-vector form as
\begin{equation}
\mathcal{A}~\boldsymbol{u}=\boldsymbol{q}~,
    \label{eq:Au=q}
\end{equation}
where $\mathcal{A}$ is the coefficient matrix (see Eq. \ref{eq:A27}), $\boldsymbol{q}$ is the source term (see Eq. \ref{eq:q}) and $\boldsymbol{u}$ is the unknown vector containing all $u_{i,j,k}$ in the interior of the computational domain. The coefficient matrix $\mathcal{A}$ has a 27-diagonal structure for a simple domain without immersed boundary. It is worth noting that the matrix $\mathcal{A}$ is not symmetric in general. The only situation where the matrix $\mathcal{A}$ is symmetric is for a domain with uniform grid spacing.

The discretization of Eq. (\ref{eq:poisson}) for all regular and irregular grid points inside the computation domain with an immersed boundary leads to a linear system 
\begin{equation}
\mathcal{B}~\boldsymbol{u}=\boldsymbol{q}+\boldsymbol{b}~,
    \label{eq:Bu=q}
\end{equation}
where $\boldsymbol{b}$ includes the known function values at the points of intersection with the immersed boundary (see Eqs. \ref{eq:bc} and \ref{eq:qb_case2}). The coefficient matrix $\mathcal{B}$ has the same dimension as matrix $\mathcal{A}$, i.e. $N\times N$ where $N=(nx-2)\times (ny-2)\times (nz-2)$. However, as discussed in the previous sections, the finite-difference schemes are modified when the 27-point stencil intersects with the immersed boundary. As a result, this introduces additional points to the 27-point compact discretization that are solution dependent, so that the coefficient matrix $\mathcal{B}$ is no longer 27-diagonal. The matrix $\mathcal{B}$ is non-symmetric even for a domain with uniform grid spacing.

\subsection{Spectral analysis: eigenvalue spectra of the coefficient matrices} \label{sec:spec}
The eigenvalue spectra of the resulting coefficient matrices for a simple domain and a domain with an immersed boundary is investigated in this section. The numerical stability involving a finite difference approximation for Eq. (\ref{eq:poisson}) requires that all eigenvalues ($\lambda$) of $\mathcal{A}$ (or $\mathcal{B}$ in the case with immersed boundaries) satisfies $real(\lambda)<0$ \citep{ZHONG2007}. The eigenvalues of the discrete finite-difference operator will depend upon the geometry and grid resolution of the problem. Therefore, it is necessary to evaluate the effect of the extended stencil due to the presence of an immersed boundary on the eigenvalue spectra of matrix $\mathcal{B}$. The eigenvalue spectra of the fourth-order-accurate compact difference scheme for two different cases are investigated on a cuboid of dimensions $[0,1]^3$: (i) A simple domain and (ii) a domain with an immersed sphere located at the center of the domain with radius 0.2.

\begin{figure}[htb!]
\centerline{
\includegraphics[width=0.8\textwidth]{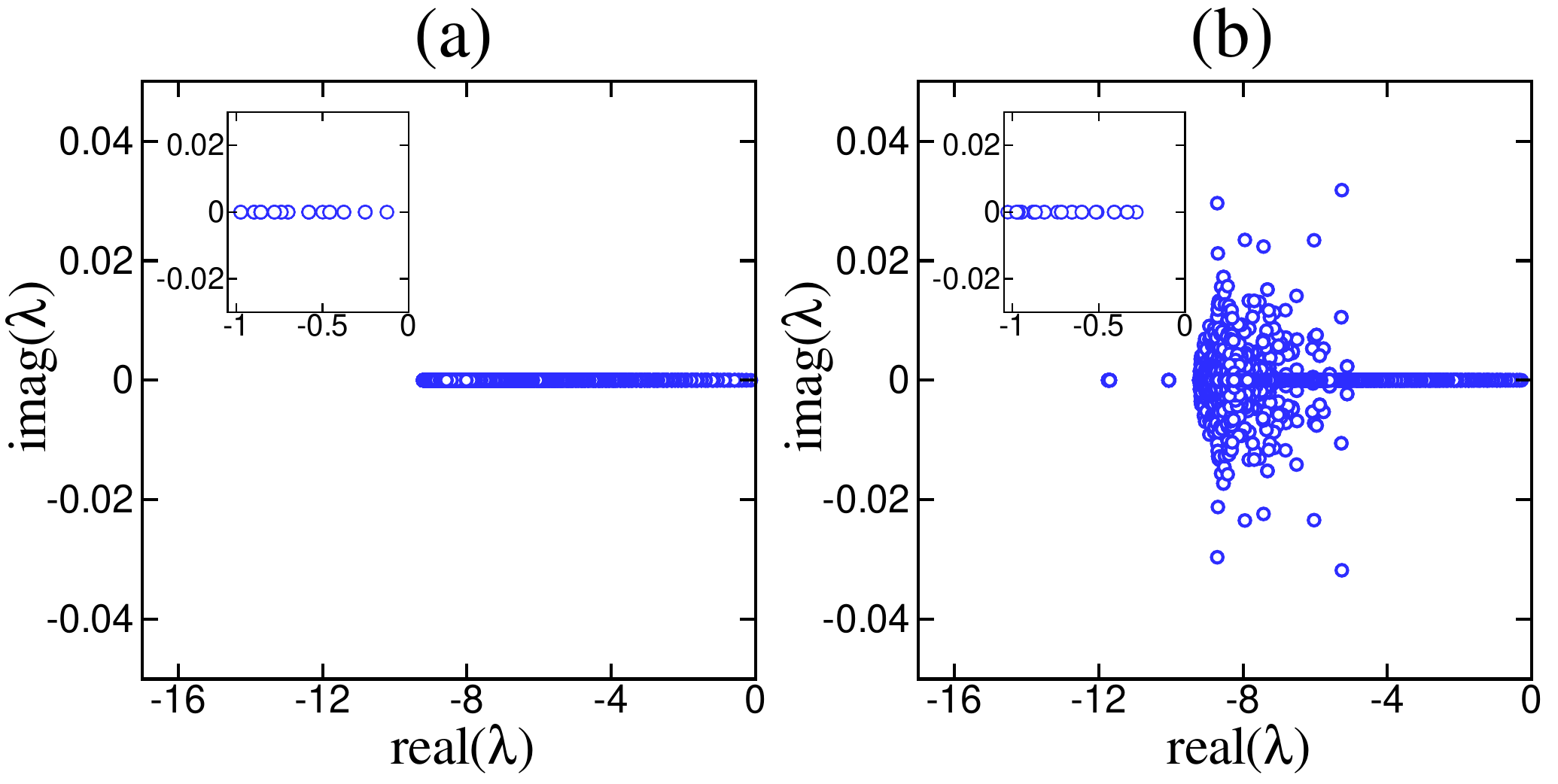}}
\caption{Eigenvalue spectra of the discretization matrices obtained for a uniform grid with $21^3$ points. (\textit{a}) Simple domain and (\textit{b}) domain with an immersed sphere with radius of 0.2. Inserts are close-up views of the spectra near the origin.}
\label{fig:spectral}
\end{figure}

The eigenvalue spectra of the resulting discretization matrix for the two cases when computed for a uniform $21^3$ grid are presented in Fig. \ref{fig:spectral}. The eigenvalues are normalized with $dx^2$. For the simple domain, all eigenvalues are real and negative as the coefficient matrix is symmetric. As can be seen the immersed boundary introduces complex eigenvalues, however, all the eigenvalues have negative real components and therefore satisfy the stability condition. 

\subsection{Preconditioned BiCGSTAB}
The fourth-order compact finite-difference discretization of Eq. (\ref{eq:poisson}) at all grid points forms a large sparse linear system. In this work, a stabilized biconjugate gradient (BiCGSTAB) method is implemented for solving the linear system, which is efficiently adjusted to account for the sharp immersed boundaries \citep{23}. Algorithm \ref{bicgstab} shows a schematic of the BiCGSTAB iterative method for a simple domain (without immersed boundary) with starting guess $\boldsymbol{u}_0$. 

As discussed in the previous section, in the presence of immersed boundaries, the coefficient matrix $\mathcal{A}$ is replaced with the matrix $\mathcal{B}$. The matrix $\mathcal{B}$ can be decomposed into
\begin{equation}
\mathcal{B}=\overline{\mathcal{A}}+\mathcal{C},
    \label{eq:Bdec}
\end{equation}
where the modified matrix $\overline{\mathcal{A}}$ has the same structure as matrix $\mathcal{A}$ (27-diagonal matrix). In fact, all the rows corresponding to the regular grid points are the same as the ones in matrix $\mathcal{A}$, however, some of the coefficients in the rows corresponding to the irregular grid points are modified. In Eq. (\ref{eq:Bdec}), the matrix $\mathcal{C}$ represents irregular entries caused by the extended stencils at the irregular grid points, see Eqs. (\ref{eq:cor}) and (\ref{eq:cor_case2}). The following changes are incorporated into algorithm \ref{bicgstab}. We have three matrix vector multiplications in algorithm \ref{bicgstab} , lines 1, 5 and 9. Thus, they are modified accordingly \vskip 2mm
\noindent $~~1:~~\boldsymbol{r}_0=\boldsymbol{q}-\overline{\mathcal{A}}\boldsymbol{u}_0+\big<\boldsymbol{b}-\mathcal{C}\boldsymbol{u}_0\big>$ \vskip 2mm
\noindent $~~5:~~~\hspace{0.066cm}\boldsymbol{v}=\overline{\mathcal{A}}\boldsymbol{\hat{p}}+\big<\mathcal{C}\boldsymbol{\hat{p}}\big>$ \vskip 2mm
\noindent $~~9:~~~\hspace{0.105cm}\boldsymbol{t}=\overline{\mathcal{A}}\boldsymbol{\hat{s}}+\big<\mathcal{C}\boldsymbol{\hat{s}}\big>$ \vskip 2mm
\noindent In the modified lines 1, 5 and 9 of algorithm \ref{bicgstab}, the $<~>$ corresponds to the additional operation due to the presence of immersed boundaries. These operations are carried out only for irregular grid points. 

\begin{algorithm}
    \caption{Preconditioned BiCGSTAB}
    \label{bicgstab}
    \begin{algorithmic}[1]
        \State $\boldsymbol{r}_0\hspace{0.04cm}=\boldsymbol{q}-\mathcal{A}\boldsymbol{u}_0$\;
        \State $\boldsymbol{p}_0=\boldsymbol{r}_0$\;
        \State \textbf{for} $i:0,1,\cdots$ \textbf{do}\;
        \State ~~~~~$\mathcal{K}\boldsymbol{\hat{p}}=\boldsymbol{p}_i$\;
        \State ~~~~~~~~~$\boldsymbol{v}=\mathcal{A}\boldsymbol{\hat{p}}$\;
        \State ~~~~~~~~\hspace{0.05cm}$\alpha=(\boldsymbol{r}_0,\boldsymbol{r}_i)/(\boldsymbol{r}_0,\boldsymbol{v})$\;
        \State ~~~~~~~~~\hspace{0.01cm}$\boldsymbol{s}=\boldsymbol{r}_i-\alpha\boldsymbol{v}$\;
        \State ~~~~~\hspace{0.05cm}$\mathcal{K}\boldsymbol{\hat{s}}=\boldsymbol{s}$\;
        \State ~~~~~~~~~\hspace{0.03cm}$\boldsymbol{t}=\mathcal{A}\boldsymbol{\hat{s}}$\;
        \State ~~~~~~~\hspace{0.1cm}$\omega=(\boldsymbol{t},\boldsymbol{s})/(\boldsymbol{t},\boldsymbol{t})$\;
        \State ~~~~\hspace{0.055cm}$\boldsymbol{u}_{i+1}=\boldsymbol{u}_i+\alpha\boldsymbol{p}+\omega\boldsymbol{s}$\;
        \State ~~~~\hspace{0.1cm}$\boldsymbol{r}_{i+1}=\boldsymbol{s}-\omega\boldsymbol{t}$\;        
        \State ~~~~~~~~~\hspace{0.01cm}$\beta=(\alpha/\omega)(\boldsymbol{r}_0,\boldsymbol{r}_{i+1})/(\boldsymbol{r}_0,\boldsymbol{r}_i)$\;
        \State ~~~~\hspace{0.055cm}$\boldsymbol{p}_{i+1}=\boldsymbol{r}_{i+1}+\beta(\boldsymbol{p}_i-\omega\boldsymbol{v})$\;
        \State \textbf{end for}
    \end{algorithmic}
\end{algorithm}

\subsection{Preconditioner based on second-order finite-difference approximation} \label{sec:precond}
Although BiCGSTAB is one of the most effective Krylov subspace methods, it can suffer from slow rate of convergence or even total breakdown. Preconditioning is a key ingredient for the improvement of both the efficiency and the robustness of Krylov subspace methods. In this section, the development of a preconditioning technique whose efficiency and cost are independent of the presence or not of an immersed boundary is presented. In algorithm \ref{bicgstab}, $\mathcal{K}$ is a preconditioner matrix which approximates $\mathcal{A}$ in some sense (or $\overline{\mathcal{A}}$ in the case with immersed boundaries). From a practical point of view, an efficient preconditioner should be low-cost to construct and the preconditioned system $\mathcal{K}\boldsymbol{\phi}=\boldsymbol{\rho}$ should be inexpensive to solve (see lines 4 and 8 in algorithm \ref{bicgstab}).

\begin{figure}[htb!]
    \centerline{
    \includegraphics[width=0.6\textwidth]{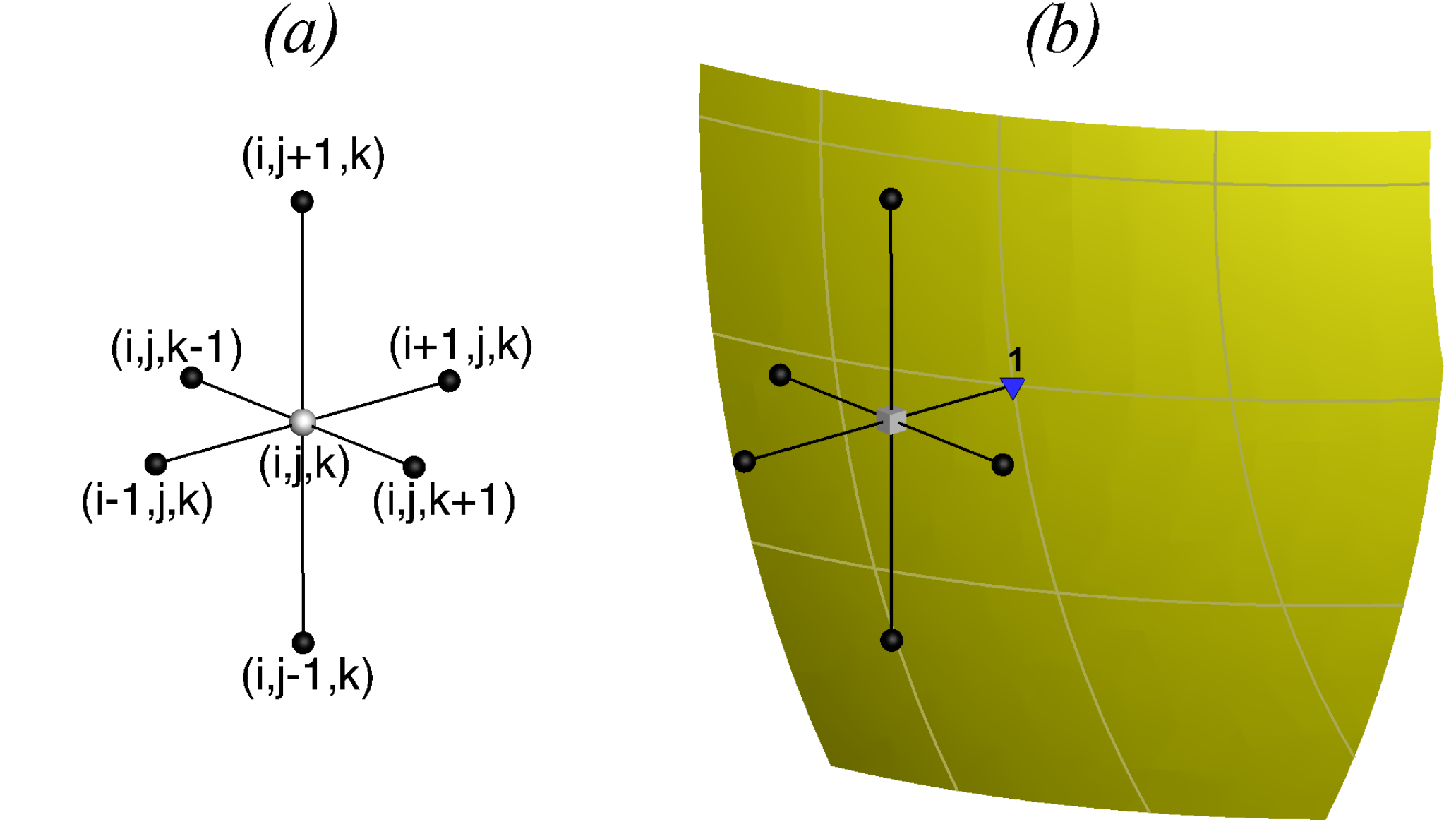}}
    \caption{(\textit{a}) Finite difference seven-point stencil at regular grid point centered at ($i,j,k$). (\textit{b}) The intersection of seven-point stencil with an immersed boundary.}
    \label{fig:stencil_2ndorder}
\end{figure}

The matrix $\mathcal{K}$ is built based on the discretization of $\nabla^2\phi=\rho$ with zero Dirichlet boundary condition on the boundaries. The second partial derivatives are approximated using a second-order standard (non-compact) finite-difference schemes in $x$, $y$ and $z$:
\begin{equation}
    \begin{alignedat}{2}
    \phi_{{xx}_{i,j,k}} &=a^x_i~\phi_{i-1,j,k} &+~b^x_i~\phi_{i,j,k} &+c^x_i~\phi_{i+1,j,k}~,\\
    \phi_{{xx}_{i,j,k}} &=a^y_j~\phi_{i,j-1,k} &+~b^y_j~\phi_{i,j,k} &+c^y_j~\phi_{i,j+1,k}~,\\
        \phi_{{zz}_{i,j,k}} &=a^z_k~\phi_{i,j,k-1} &+~b^z_k~\phi_{i,j,k} &+c^z_k~\phi_{i,j,k+1}~,
    \end{alignedat}
    \label{eq:2ndorder}
\end{equation}
where the coefficients in the $x-$direction are given by
\begin{equation}
a^x_i=\frac{2}{dx_b(dx_f+dx_b)}~, ~~~ b^x_i=-\frac{2}{dx_b dx_f}~, ~~~ c^x_i=\frac{2}{dx_f(dx_f+dx_b)}~,
    \label{eq:2ndorder_coef}
\end{equation}
where $dx_f=x_{i+1}-x_i$ and $dx_b=x_i-x_{i-1}$. The coefficients in $y-$ and $z-$directions in Eq. (\ref{eq:2ndorder}) are the same except that we replace $x$ with $y$ and $z$, respectively. Incorporating Eq. (\ref{eq:2ndorder}) into $\nabla^2\phi=\rho$ leads to a seven-point, second-order scheme at the regular grid point ($i,j,k$) as shown in Fig. \ref{fig:stencil_2ndorder}(a)
\begin{equation}
a^z_k~\phi_{i,j,k-1}+a^y_j~\phi_{i,j-1,k}+a^x_i~\phi_{i-1,j,k}+(b^z_k+b^y_j+b^x_i)~\phi_{i,j,k}+c^z_k~\phi_{i,j,k+1}+c^y_j~\phi_{i,j+1,k}+c^x_i~\phi_{i+1,j,k}=\rho_{i,j,k}~.
    \label{eq:A7}
\end{equation}

Now consider the seven-point stencil centered at the irregular grid point located at ($i,j,k$) which intersects with an immersed boundary as illustrated in Fig. \ref{fig:stencil_2ndorder}(b). We use the same irregular grid point and immersed boundary as shown in Fig. \ref{fig:irr_grid_point}. The immersed boundary intercepted point, $IBI_1$ is located at $x_i<x_{{IBI}_1}<x_{i+1}$. As a result, the second derivative in the $x-$direction needs to be adjusted to take into account the immersed boundary. Therefore
\begin{equation}
    \phi_{{xx}_{i,j,k}} =\overline{a^x_i}~\phi_{i-1,j,k} +\overline{b^x_i}~\phi_{i,j,k} +\overline{c^x_i}~\phi_{{IBI}_1}~,
    \label{eq:2ndorder_xx}
\end{equation}
with the modified coefficients 
\begin{equation}
\overline{a^x_i}=\frac{2}{dx_b(\overline{dx_f}+dx_b)}~, ~~~ \overline{b^x_i}=-\frac{2}{dx_b \overline{dx_f}}~, ~~~ \overline{c^x_i}=\frac{2}{\overline{dx_f}(\overline{dx_f}+dx_b)}~,
    \label{eq:2ndorder_coef_mod}
\end{equation}
where $\overline{dx_f}=x_{{IBI}_1}-x_i$. It should be noted that $\phi_{{IBI}_1}=0$. This leads to the modified stencil at the irregular grid point centered at ($i,j,k$)
\begin{equation}
a^z_k~\phi_{i,j,k-1}+a^y_j~\phi_{i,j-1,k}+\overline{a^x_i}~\phi_{i-1,j,k}+(b^z_k+b^y_j+\overline{b^x_i})~\phi_{i,j,k}+c^z_k~\phi_{i,j,k+1}+c^y_j~\phi_{i,j+1,k}+\hcancel[red]{c^x_i~\phi_{i+1,j,k}}=\rho_{i,j,k}~.
    \label{eq:A7_mod}
\end{equation}

Note that the grid point located at ($i+1,j,k$) falls into $\Omega^{-}$, hence, it is dropped in the above equation (struck through in Eq. \ref{eq:A7_mod}). It is important to note that the preconditioner matrix $\mathcal{K}$ has a seven-diagonal structure 
with or without the presence of an immersed boundary. It is thus a convenient algorithm that easily accommodates immersed boundaries, yet the cost to build and solve the preconditioned system is independent of the complexity of the geometry and the presence or not of an immersed boundary. In the present work, this preconditioned system is solved iteratively using the modified strongly implicit (MSI) procedure \citep{24}. The MSI method has very good convergence properties was shown to outperform Stone's strongly implicit procedure \citep{stone}, incomplete lower-upper decomposition (ILU) and successive over-relaxation
methods. 

\subsection{Floating-point operation counts} \label{sec:flop}
Another important aspect of the solution of Poisson equation with a high-order discretization method is the extra floating-point operation (FLOP) of the algorithm components due to the presence of an immersed body. The preconditioned BiCGSTAB algorithm is built from several basic components: Sparse matrix vector products (SPMV) such as $\mathcal{A}\boldsymbol{\hat{v}}$, inner products of two vectors (DOT-PROD) like $(\boldsymbol{r}_0,\boldsymbol{r}_i)$ and combined scalar vector multiplication and vector addition (SVPV) for example $\boldsymbol{s}-\omega\boldsymbol{t}$. Moreover, algorithm 1 contains two preconditioned systems (PCS) that need to be solved. Table \ref{tab:flop} lists the FLOP counts for the preconditioned BiCGSTAB algorithm per outer iteration for a simple domain and a domain with immersed boundary.

In Table \ref{tab:flop}, $iter_{prec}$ refers to the number of inner iteration for solving the preconditioned system, $N$ is the total number of grid points and $N_{irr}$ is the number of irregular grid points. Therefore, the extra FLOP counts due to the presence of an immersed body as a percentage of FLOPs for a simple domain is
\begin{equation}
FLOP_{extra}(\%)=\frac{N_{irr}}{N}\frac{432}{126+80\times iter_{prec}}\times100~.
    \label{eq:flop}
\end{equation}

It should be noted that the FLOP counts for $<~>$, which corresponds to the operations carried out for irregular grid points come from the fact that there are nine FD operators for each direction and for each FD operator, there is a maximum of four additional grid points, resulting in 8 FLOPs (see Eq. \ref{eq:cor} for example). Therefore, the maximum FLOP count for each irregular grid point will be $3\times9\times8=216$ FLOPs. For $iter_{prec}=4$ which is typical for 
\begin{table}[hbt!]
\centerline{
\begin{tabular}{lcc} \hline
                 & simple domain  & domain with an immersed boundary \Tstrut\Bstrut \\ \hline
SPMVS            & 106$N$                  & 106$N$+2$\times$ 216$N_{irr}$      \Tstrut \\
DOT-PROD~\&~SVPV & 20$N$                   & 20$N$                   \\
PCS              & 2$\times$ 40$N\times iter_{prec}$ & 2$\times$ 40$N\times iter_{prec}$ \\
total FLOP counts& $(126+80\times iter_{prec})N$ & $(126+80\times iter_{prec})N + 432N_{irr}$ \\ \hline
\end{tabular}}
\caption{FLOP counts for main components of algorithm \ref{bicgstab} for domains without and with immersed boundaries.}
\label{tab:flop}
\end{table}
a preconditioned BiCGSTAB solver, we could have $FLOP_{extra}(\%)\approx N_{irr}/N\times100$ which in most cases is negligible. It is worth mentioning that a different iterative solver for the preconditioned system ($\mathcal{K}\boldsymbol{\phi}=\boldsymbol{\rho}$) will result in a different FLOP count (PCS). However, the number of FLOPs for PCS will be the same regardless of the presence or absence of an immersed body.

\section{Numerical experiments} \label{sec:results}
In this section, numerical experiments are conducted to demonstrate the accuracy and efficiency of the present method for solving Eq. (\ref{eq:poisson}) on uniform and nonuniform grids with immersed boundaries. All problems are solved on the cuboid domain $\Omega$ with the same number of grid points in the $x$, $y$ and $z$ directions. The order of accuracy is computed using the definition
\begin{equation}
    p=log\bigg(\frac{||E_{n_1}||_\infty}{||E_{n_2}||_\infty}\bigg)~\Bigg/~log\bigg(\frac{n_2-1}{n_1-1}\bigg)  ~,
    \label{eq:order}
\end{equation}
where $||E_{n_1}||_\infty$ and $||E_{n_2}||_\infty$ are the infinity norms of the errors between the analytical, $u^{(ex)}$, and the numerical, $u$, solutions for two different grids with $n_1$ and $n_2$ points in each direction. For the preconditioner solver, eight inner iterations are performed for all grid sizes. The initial guess is the zero vector.

\subsection{Problem 1} \label{sec:p1}
\begin{figure}[hbt!]
    \centerline{
    \includegraphics[width=0.8\textwidth]{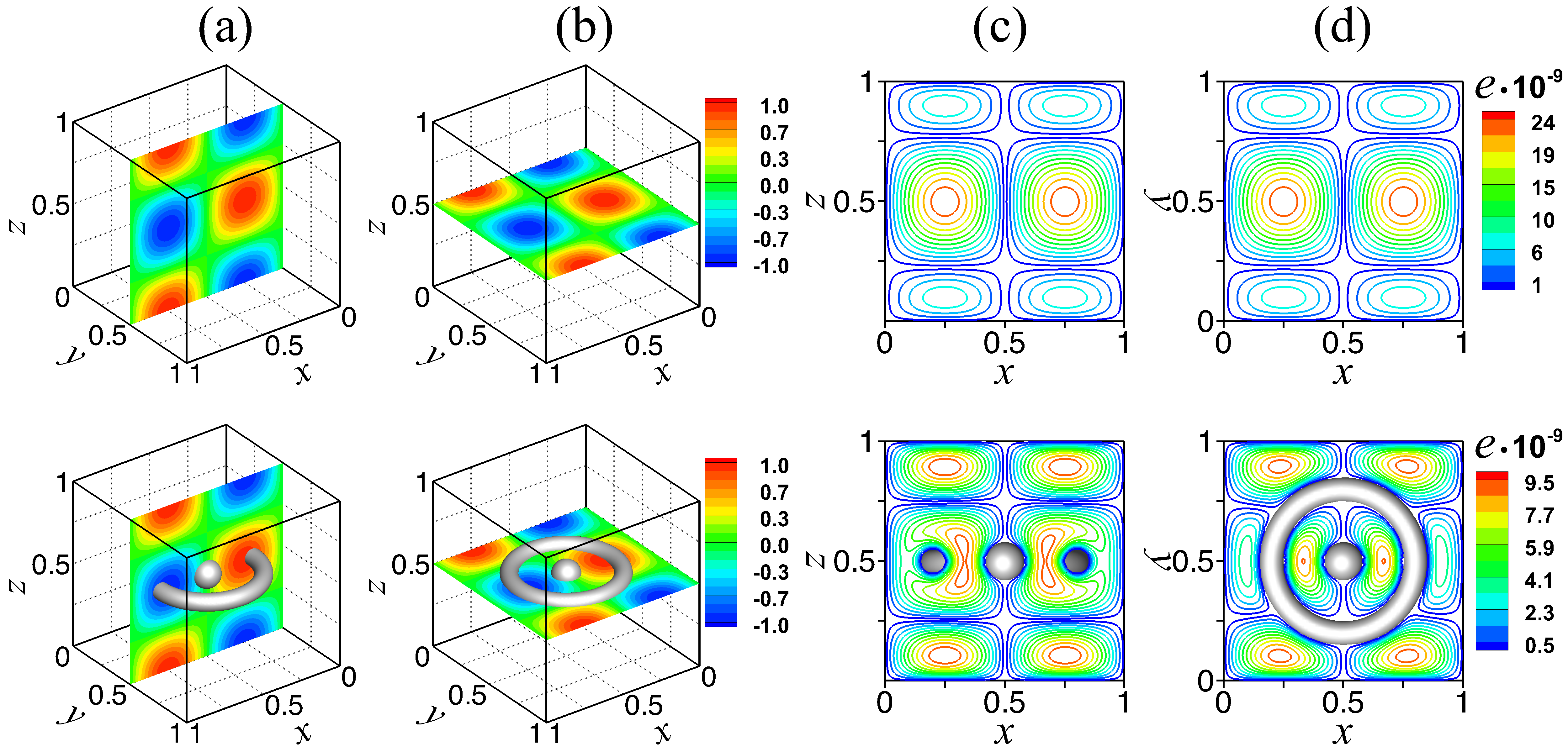}}
    \caption{Numerical solutions and local errors of Eq. (\ref{eq:p1}) on the computational domain with $129^3$ uniform grid points for two different cases. Simple domain without immersed boundary (top) and domain with an immersed toroidal surface plus an immersed sphere (bottom). Contours of numerical solutions at (\textit{a}) $y=0.5$ and (\textit{b}) $z=0.5$. Contour lines of errors correspond to the numerical solutions at (\textit{c}) $y=0.5$ and (\textit{d}) $z=0.5$.}
    \label{fig:p1_sol_err}
\end{figure}

\begin{figure}[htb!]
\centerline{
\noindent
\begin{minipage}{.4\textwidth}
\quad\quad\quad\quad\quad\quad\quad\quad\quad\quad~(a)
\end{minipage}
\begin{minipage}{.4\textwidth}
\quad\quad\quad\quad\quad\quad\quad\quad\quad~~~(b)
\end{minipage}}
    \centerline{
    \includegraphics[width=0.8\textwidth]{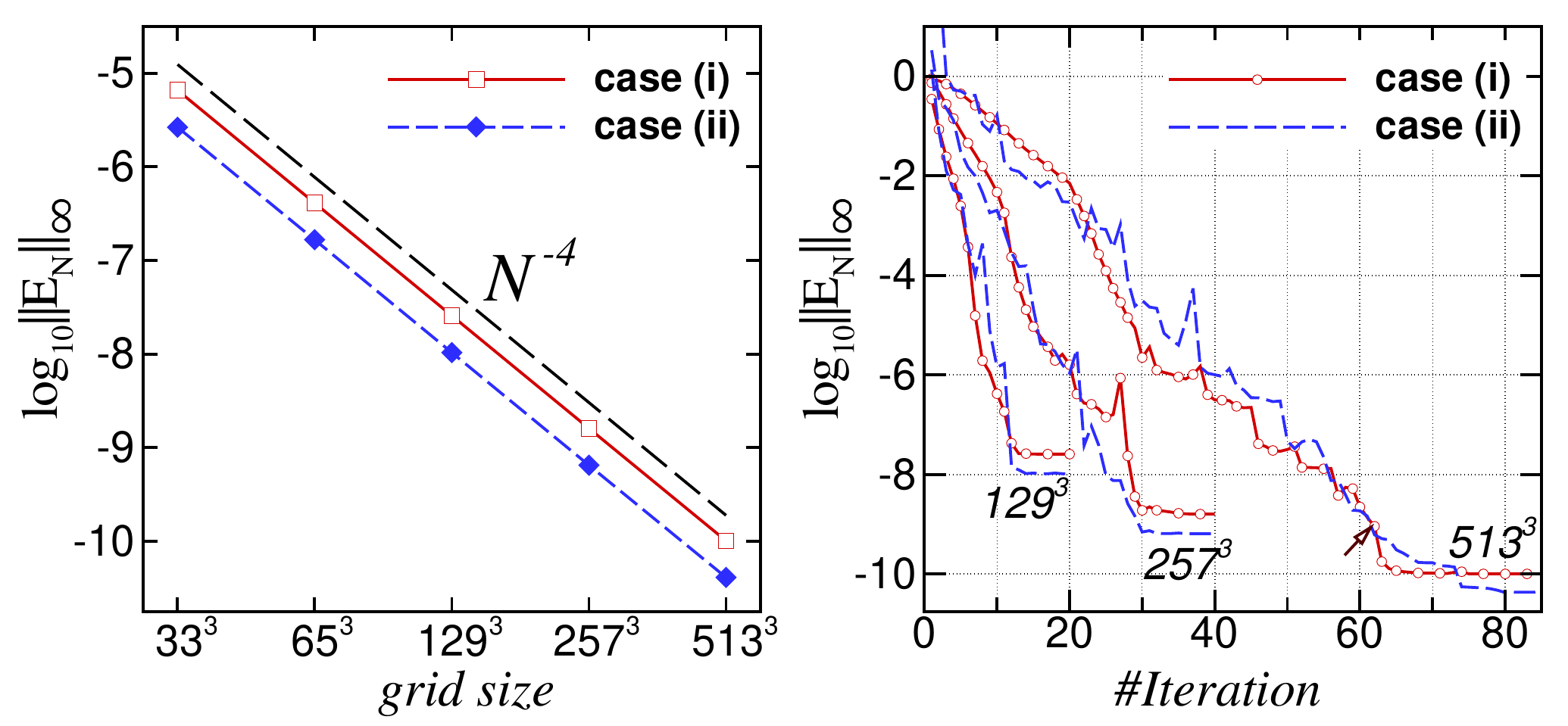}}
    \caption{(\textit{a}) Numerical error in the infinity-norm as a function of grid points for solutions of Eq. (\ref{eq:p1}) for domains without and with immersed boundaries. (\textit{b}) Convergence behaviour of the numerical error versus outer iteration number for various grid sizes for the domain with and without the immersed boundaries.}
    \label{fig:p1_err_resid}
\end{figure}

To demonstrate the accuracy of the new method and the efficiency of the preconditioned BiCGSTAB solver, solving the following Poisson equation with Dirichlet boundary conditions is considered:
\begin{equation}
\begin{aligned}
    u_{xx}+u_{yy}+u_{zz}&=-3\omega^2~sin(\omega x)~cos(\omega y)~cos(\omega z)~, & \text{in}~~\Omega^{+}~, \\
    u(x,y,z) &=sin(\omega x)~cos(\omega y)~cos(\omega z)~, \quad\quad & \text{on}~~\Gamma~\&~\partial\Omega~,
\end{aligned}
\label{eq:p1}
\end{equation}
where $\omega=2\pi$. Eq. (\ref{eq:p1}) is solved on a cuboid of dimensions $[0,1]^3$ with a uniform grid in the $x$, $y$ and $z$ directions for two cases: (i) simple domain without immersed boundary and (ii) a domain with one immersed toroidal surface with an outer radius of 0.35 and an inner radius of 0.05 plus an immersed sphere with radius 0.08 (see Fig. \ref{fig:p1_sol_err}, bottom plots). Computed solutions and the corresponding errors, $e=|u^{(ex)}-u|$ are shown in Fig. \ref{fig:p1_sol_err} on a $129^3$ grid. The reader's attention is drawn to the sharp interface of the obtained solution, and the relative smoothness of the error distribution in the vicinity of the immersed boundary. The solution and the corresponding error is computed for several grids. The errors in the infinity-norm, $||E_N||_\infty$, are plotted as a function of grid size in Fig. \ref{fig:p1_err_resid}(a). The results confirm
the fourth-order accuracy of the new method for both the simple domain and, most importantly, also for the domain with immersed boundaries. No loss of accuracy for the domain with immersed boundary is observed.

In the following, the convergence characteristics and the $FLOP_{extra}$ of the new method for solving Eq. (\ref{eq:p1}) is compared between the simple domain and domain with immersed bodies. Figure \ref{fig:p1_err_resid}(b) displays the convergence rates for various grid sizes for the two cases described above. It shows the infinity-norm of the error after each outer iteration. The convergence behaviour is very similar for the two cases for the different grid sizes. With a proper design of the discretization operator and the preconditioning strategy, the new solution strategy is proven to be equally efficient for domains with immersed boundaries and for simple domains (without immersed boundaries). Table \ref{tab:p1} presents the extra FLOP counts for the problem with immersed boundaries. For the problem sizes of interest, Table \ref{tab:p1} shows that the $FLOP_{extra}$ due to the presence of the immersed boundary is negligible for the preconditioned BiCGSTAB solver. For the largest grid size investigated, the $FLOP_{extra}$ is about 0.11\%.

\begin{table}[htb!]
\captionsetup{width=0.655\textwidth}
    \caption{Comparison of extra FLOP for the second case with immersed boundaries using the preconditioned BiCGSTAB method.}
    \centerline{
    \begin{tabular}{cccccc} \hline
    Grid size $\rightarrow$  & $33^3$ & $65^3$ & $129^3$ & $257^3$ & $513^3$ \Tstrut\Bstrut \\ \hline
      $N_{irr}$     &~~~1370~~~&~~~5018~~~&~~17682~~~&~~66322~~~&~~261634~~ \Tstrut\\ 
 $FLOP_{extra}(\%)$ &  2.15    &   1.03   &  0.46    & 0.22     & 0.11 \Bstrut\\ \hline
    \end{tabular}}
    \label{tab:p1}
\end{table}

\subsection{Problem 2} \label{sec:p2}
\begin{figure}[htb!]
    \centerline{
    \includegraphics[width=0.8\textwidth]{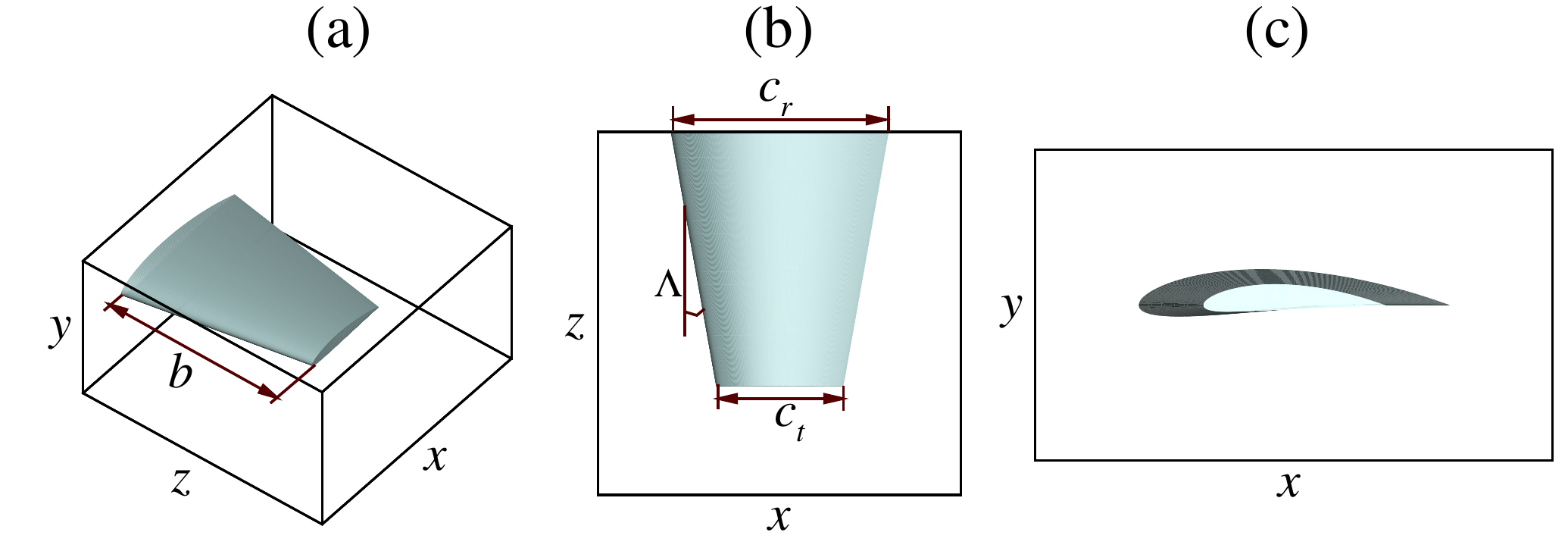}}
    \caption{Finite tapered wing used as an immersed body in Eq. (\ref{eq:p2}). Perspective view (\textit{a}), planform view (\textit{b}) and side view (\textit{c}). Shown are the wing span $b$, wing tip chord $c_t$, wing root chord length $c_r$ and wing sweep angle $\Lambda=atan[(c_r-c_t)/2b]$.}
    \label{fig:p2_setup}
\end{figure}

Consider the following equation with Dirichlet boundary conditions:
\begin{equation}
\begin{aligned}
    u_{xx}+u_{yy}+u_{zz}&=(\sigma^2-\omega^2)~sin(\omega x)~sin(\omega z)~e^{\sigma y}, & \text{in}~~\Omega^{+}~, \\
    u(x,y,z) &=sin(\omega x)~sin(\omega z)~e^{\sigma y}~, \quad\quad & \text{on}~~\Gamma~\&~\partial\Omega~,
\end{aligned}
\label{eq:p2}
\end{equation}
with $\omega=2\pi$ and $\sigma=-10$. In this example, the computational domain is set to $[0,1]\times[−0.3,0.3]\times[0,1]$ and the immersed body is a finite tapered wing with a NACA 5514 airfoil section. Figure \ref{fig:p2_setup} illustrates the parameters that define the wing shape which are the wing span $b=0.7$, the wing root chord $c_r=0.6$, and the tip chord $c_t=0.35$. Numerical investigations of the external flow over a finite wing represent one of the difficult challenges in the field of computational fluid dynamics. One of the main difficulties lies in the generation of a high-quality mesh around the wing tip. Sharp immerse interface method are therefore very useful for that type of problems.

Eq. (\ref{eq:p2}) is solved on a uniform grid in the $x$ and $z$ directions while a non-uniform grid is used in the $y$ direction, for which the stretching function is given by
\begin{equation}
y_j=-0.35+L\bigg\{1+\frac{sinh[\beta(\eta-s)]}{sinh(\beta s)}\bigg\},\quad  ~\eta=\frac{j-1}{ny-1}, \quad ~1\leq j\leq ny~. \label{eq:p2_y}
\end{equation}
The parameter $s$ is defined as
\begin{equation}
s=\frac{1}{2\beta}log\bigg[\frac{1+(e^{+\beta}-1)(L/H)}{1+(e^{-\beta}-1)(L/H)}\bigg]~, \label{eq:p2_s}
\end{equation}
where $L=0.3$ and $H=0.6$. The parameter $\beta$ is the clustering parameter which controls the nonuniformity of the grid around $y=0$. It is set to $\beta=3$ which puts more grid points around $y=0$.

\begin{figure}[hbt!]
    \centerline{
    \includegraphics[width=0.8\textwidth]{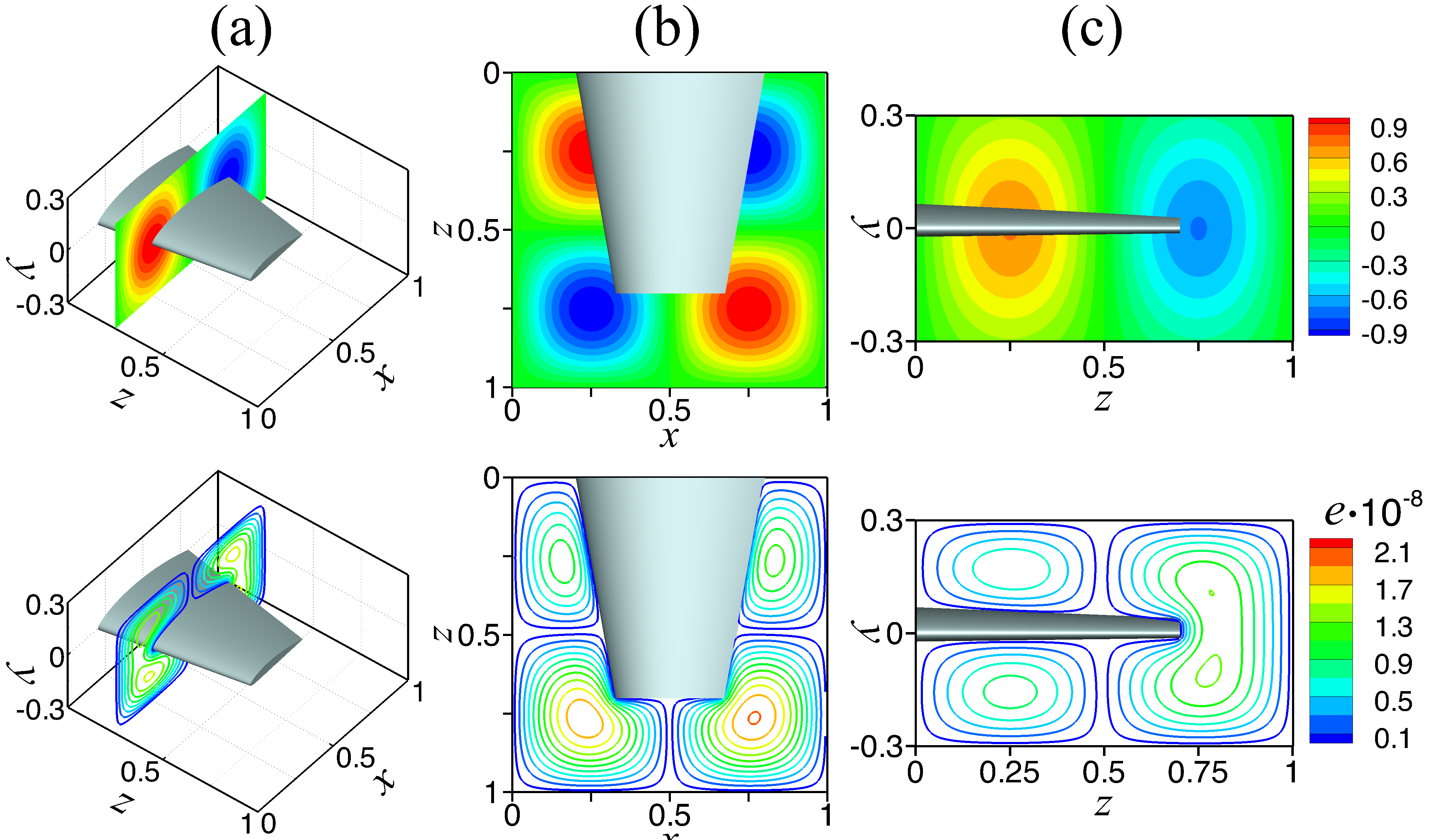}}
    \caption{Contours of numerical solution (top) and contour lines of the corresponding error (bottom) computed for Eq. (\ref{eq:p2}) on the computational domain with $129^3$ grid points. (\textit{a}) plane $z=0.25$, (\textit{b}) plane $y=0$ and (\textit{c}) plane $x=0.4$.}
    \label{fig:p2}
\end{figure}

Figure \ref{fig:p2} displays the numerical solution to Eq. (\ref{eq:p2}) and the corresponding error for a grid with $129^3$ points. The solutions are sharp across the interfaces and the local error has a smooth distribution in the vicinity of the immersed boundary. Table \ref{tab:p2} shows the numerical errors based on the infinity-norm of the current method for different sets of grids which demonstrates that the method is fourth-order-accurate. The extra FLOP as a result of the immersed
body in this problem is presented in Table \ref{tab:p2_flop}, which varies from 0.11(\%) to 1.68(\%) depending on the grid size.

\begin{table}[htb!]
\captionsetup{width=0.46\textwidth}
\centerline{
\begin{tabular}{cccc} \hline
 Grid size & $||E_N||_{\infty}$      & Error ratio & Order \Tstrut\Bstrut \\ \hline
 $33^3$    & $5.5738\times10^{-06}$  &          &           \Tstrut\\ [0.2em]
 $65^3$    & $3.4225\times10^{-07}$  & ~16.286~ &  ~4.026~  \\ [0.2em]
 $129^3$   & $2.1196\times10^{-08}$  & ~16.147~ &  ~4.013~  \\ [0.2em]
 $257^3$   & $1.3177\times10^{-09}$  & ~16.086~ &  ~4.008~  \\ [0.2em]
 $513^3$   & $8.1693\times10^{-11}$  & ~16.130~ &  ~4.012~  \\
\hline
\end{tabular}}
\caption{Maximum absolute error and order of accuracy of the current method for test problem \ref{sec:p2}.}
\label{tab:p2}
\end{table}

\begin{table}[htb!]
\captionsetup{width=0.66\textwidth}
    \centerline{
    \begin{tabular}{cccccc} \hline
    Grid size $\rightarrow$  & $33^3$ & $65^3$ & $129^3$ & $257^3$ & $513^3$ \Tstrut\Bstrut \\ \hline
      $N_{irr}$     &~~~1071~~~ &~~~4119~~~&~~16279~~~&~~64727~~~&~~258444~~ \Tstrut\\
 $FLOP_{extra}(\%)$ &  1.68     &  0.85    &  0.43    &  0.22    &  0.11     \\ \hline
    \end{tabular}}
    \caption{Comparison of extra FLOP for the second case with immersed boundaries using the preconditioned BiCGSTAB method.}
    \label{tab:p2_flop}
\end{table}

\subsection{Problem 3} \label{sec:p3}
\begin{figure}[hbt!]
    \centerline{
    \includegraphics[width=0.77\textwidth]{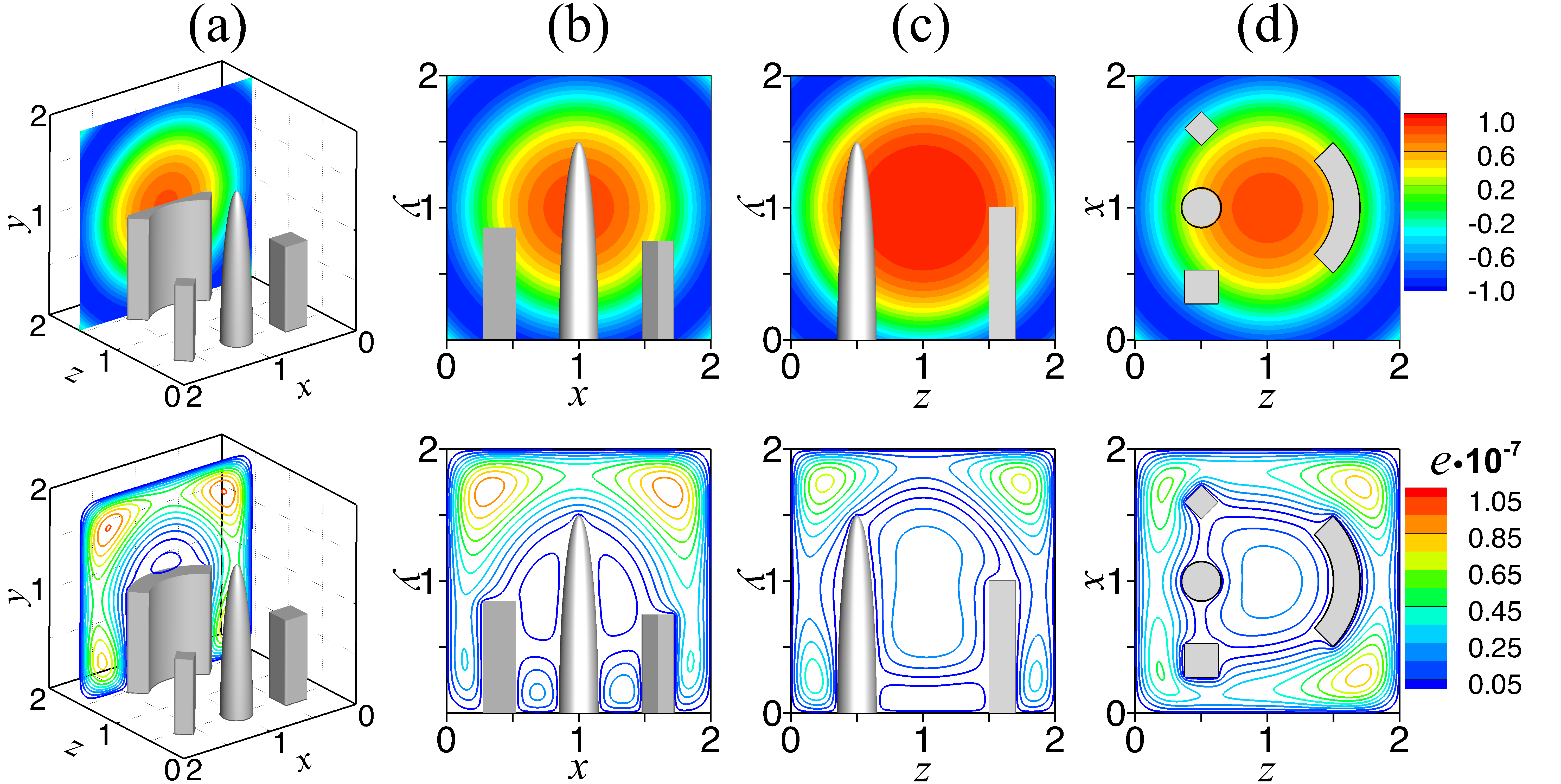}}
    \caption{Numerical solution (top) and contour lines of local error (bottom) computed for Eq. (\ref{eq:p3}) on the computational domain with $129^3$ grid points. (\textit{a}) plane $z=1.55$, (\textit{b}) plane $z=0.5$, (\textit{c}) plane $x=1$ and (\textit{d}) plane $y=0.5$.}
    \label{fig:p3_sol_err}
\end{figure}

Next, we consider solving the following Poisson equation on a uniform grid
\begin{equation}
\begin{aligned}
    u_{xx}+u_{yy}+u_{zz}&=-6\omega~sin(\omega r)-4\omega^2~r~cos(\omega r)~, & \text{in}~~\Omega^{+}~, \\
    u(x,y,z) &=cos(\omega r)~, \quad\quad & \text{on}~~\Gamma~\&~\partial\Omega~,
\end{aligned}
\label{eq:p3}
\end{equation}
where $r=(x-1)^2+(y-1)^2+(z-1)^2$ and $\omega=2$. Eq. (\ref{eq:p3}) is solved in the cuboid domain with dimensions $[0,2]^3$ in presence of four immersed bodies as illustrated in Fig. \ref{fig:p3_sol_err}. In this example, the immersed bodies imitate tall buildings encountered in wind engineering, where wind-induced load effects on buildings with different heights and proximities are regularly investigated. Fig. \ref{fig:p3_sol_err} shows the numerical solution computed by the preconditioned BiCGSTAB method using $129^3$ grid points as well as the contour lines of the local errors.
\newpage
\begin{table}[htb!]
\captionsetup{width=0.46\textwidth}
\centerline{
\begin{tabular}{cccc} \hline
 Grid size & $||E_N||_{\infty}$      & Error ratio & Order \Tstrut\Bstrut  \\ \hline
 $33^3$    & $2.7667\times10^{-05}$  &       &         \Tstrut\\ [0.2em]
 $65^3$    & $1.7340\times10^{-06}$  & ~15.955~ &  ~~3.996~  \\ [0.2em]
 $129^3$   & $1.0841\times10^{-07}$  & ~15.994~ &  ~~3.999~  \\ [0.2em]
 $257^3$   & $6.7764\times10^{-09}$  & ~15.999~ &  ~~3.999~  \\ [0.2em]
 $513^3$   & $4.2265\times10^{-10}$  & ~16.033~ &  ~~4.003~  \\
\hline
\end{tabular}}
\caption{Numerical error in the infinity-norm and order of accuracy for solutions of Eq. (\ref{eq:p3}) for a domain with immersed boundaries.}
\label{tab:p3}
\end{table}

The results of the error study are presented in Table \ref{tab:p3} which demonstrates that the method is fourth-order-accurate based on the error measured in the infinity-norm.

\subsection{Problem 4} \label{sec:p4}
\begin{figure}[htb!]
    \centerline{
    \includegraphics[width=0.6\textwidth]{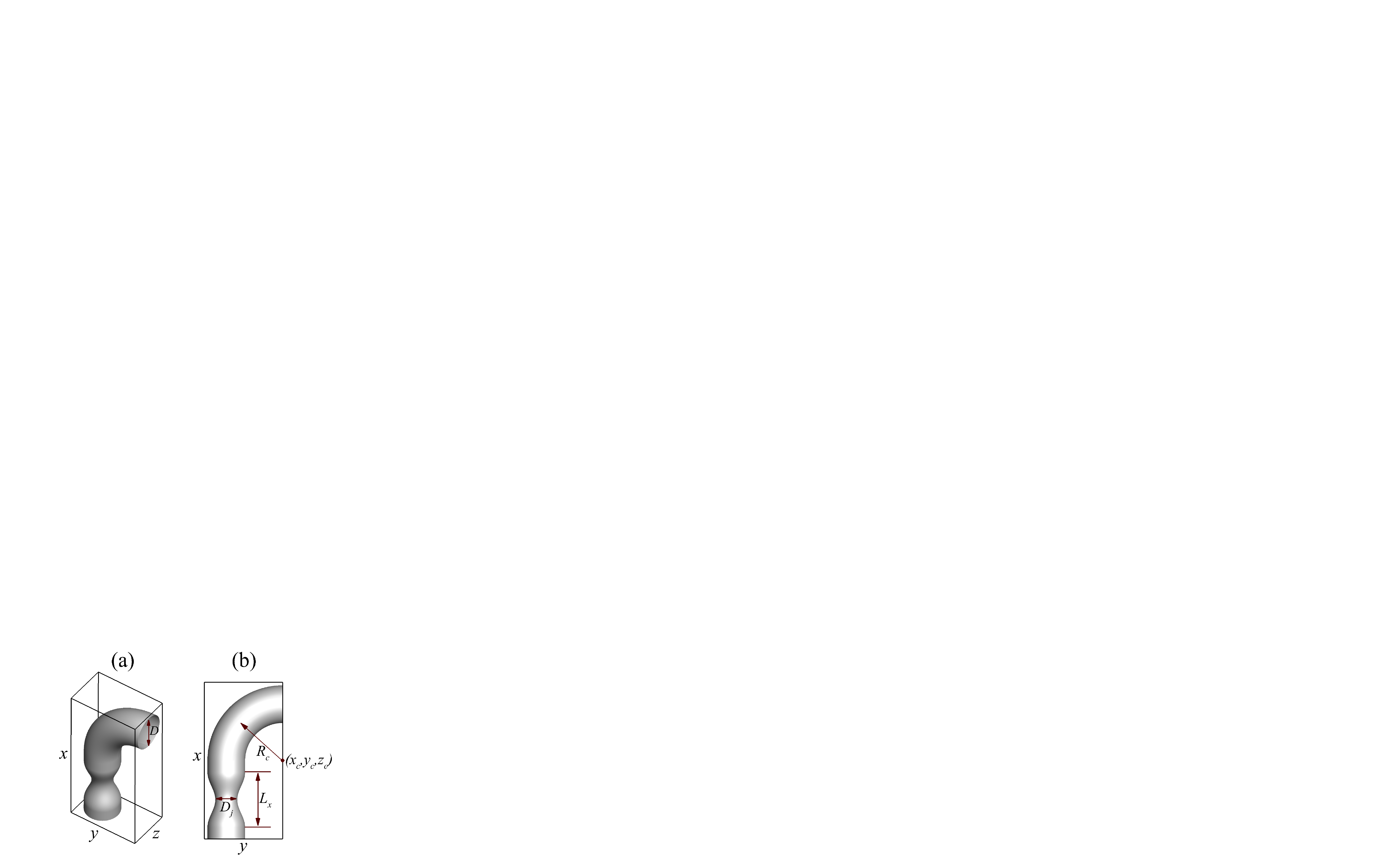}}
    \caption{Schematic of immersed boundary used in problem \ref{sec:p4}. Perspective view (\textit{a}) and side view (\textit{b}).}
    \label{fig:p4_setup}
\end{figure}

For the final test case, the following equation with Dirichlet boundary condition on a nonuniform grid with irregular boundaries is solved:
\begin{equation}
\begin{aligned}
    u_{xx}+u_{yy}+u_{zz}&=2\sigma e^{\sigma r} (3+2\sigma r)~, & \text{in}~~\Omega^{+}~, \\
    u(x,y,z) &=e^{\sigma r}~, \quad\quad & \text{on}~~\Gamma~,
\end{aligned}
\label{eq:p4}
\end{equation}
where $r=(x-0.5)^2+(y-0.5)^2+(z-0.5)^2$ and $\sigma=-10$. The domain $\Omega^{+}$ is the region inside the boundary $\Gamma$ as shown in Fig. \ref{fig:p4_setup}. In contrast to the other test problems, Eq. (\ref{eq:p4}) is solved inside the immersed boundary. This test problem is designed to show the accuracy and efficiency of the new method for simulations of internal flows such as the flow through a channel or stenosed aorta.

Grid points are distributed uniformly in the $x$ direction for $0\leq x\leq1$ while the stretching function for the nonuniform grids in the $y-$ and $z-$directions is given by
\begin{align}
y_j&=L\bigg\{1+\frac{sinh[\beta(\eta-s)]}{sinh(\beta s)}\bigg\},\quad\quad\quad  ~\eta=\frac{j-1}{ny-1}, ~~1\leq j\leq ny~, \label{eq:p4_y} \\
z_k&=L\bigg\{1+\frac{sinh[\beta(\eta-s)]}{sinh(\beta s)}\bigg\}+ 0.35,  ~~\eta=\frac{k-1}{nz-1}, ~~1\leq k\leq nz~, \label{eq:p4_z} 
\end{align}
where $\beta=2$ and $\beta=4$ for Eqs. (\ref{eq:p4_y}) and (\ref{eq:p4_z}), respectively. The parameter $s$ given by Eq. (\ref{eq:p2_s}) is computed with $L=0.14$ \& $H=0.5$ for Eq. (\ref{eq:p4_y}) and with $L=0.15$ \& $H=0.3$ for Eq. (\ref{eq:p4_z}). This test case could be a
\begin{figure}[hbt!]
    \centerline{
    \includegraphics[width=0.8\textwidth]{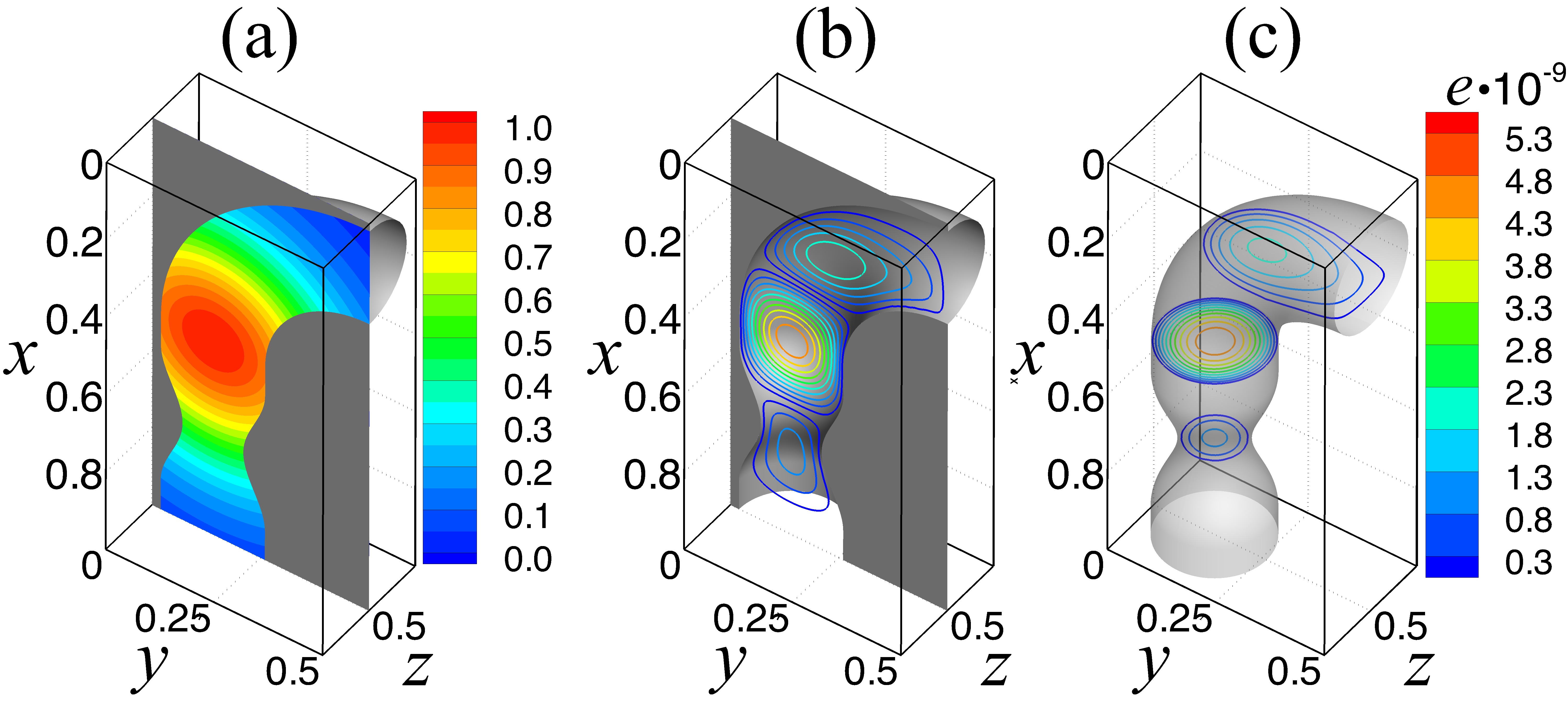}}
    \caption{Numerical solution and error for problem \ref{sec:p4} on nonuniform grid with irregular boundary. Computed solution in the plane $z=0.5$ (\textit{a}), the corresponding local error in the plane $z=0.5$ (\textit{b}) and planes $x=0.7$, $x=0.4$ and $x=0.2$ (\textit{c}).}
    \label{fig:p4_sol_err}
\end{figure}
model of stenosis pipe flow with 66\% reduction of the area. Shown in Fig. \ref{fig:p4_setup} is a curved pipe centered at $x_c=0.5$, $y_c=0.5$ and $z_c=0.5$ with an inner diameter of $D_i=0.24$. The inner diameter of the reduced area is defined as
\begin{equation}
D_j=D_i-\gamma\bigg\{1+cos\big[\frac{2\pi}{L_x}(x-x_0)\big]\bigg\}, ~~~ |x-x_0|\le L_x~,
    \label{eq:p4_d}
\end{equation}
where $\gamma=0.05$ is a coefficient to apply a 66\% area reduction, $L_x(=0.35)$ denotes the length of the stenosis and its center is located at $x_0=0.75$. The numerical solutions and error obtained for Eq. (\ref{eq:p4}) are illustrated in Fig. \ref{fig:p4_sol_err} for a grid size of $129^3$. Different grid sizes ranging from $33^3$ to $513^3$ were used to evaluate the order of accuracy summarized in Table \ref{tab:p4}. It can be observed that the new method preserves its fourth-order accuracy for this problem as well. Furthermore, the extra FLOP counts per outer iteration varies from 6.3\% for the smallest grid size to 0.46\% for the largest grid size.

\begin{table}[hb!]
\captionsetup{width=0.46\textwidth}
\centerline{
\begin{tabular}{cccc} \hline
 Grid size & $||E_N||_{\infty}$      & Error ratio & Order \Tstrut\Bstrut \\ \hline
 $33^3$    & $1.3271\times10^{-06}$  &        &         \Tstrut\\ [0.2em]
 $65^3$    & $8.3864\times10^{-08}$  & ~15.824~ &  ~~3.984~  \\ [0.2em]
 $129^3$   & $5.2478\times10^{-09}$  & ~15.981~ &  ~~3.998~  \\ [0.2em]
 $257^3$   & $3.2820\times10^{-10}$  & ~15.990~ &  ~~3.999~  \\ [0.2em]
 $513^3$   & $2.0647\times10^{-11}$  & ~15.896~ &  ~~3.991~  \\
\hline
\end{tabular}}
\caption{Maximum absolute error and order of accuracy of the current method for test problem \ref{sec:p4}.}
\label{tab:p4}
\end{table}

\section{Conclusion} \label{sec:end}
In this paper, the development of a sharp-interface and high-order method to solve the three-dimensional Poisson equation with Dirichlet boundary conditions on nonuniform grids with immersed boundaries is presented. Fourth-order compact difference schemes are used to discretize the Poisson equation on the nonuniform grid resulting in 27-point FD regular stencils. The sharp representation of the interface of the immersed body is accomplished by modifying the finite difference stencils at irregular grid points (near the immersed boundary). In particular, additional grid points are added to the standard 27-point FD stencils to obtain a sharp solution across the interface while retaining the fourth-order formal accuracy.

A preconditioned BiCGSTAB method is designed to solve the large sparse algebraic system that arose from the discretization of the Poisson equation. A second-order standard (non-compact) finite-difference scheme is employed to discretize the preconditioned system leading to a seven-diagonal matrix. The extra floating-point operation ($FLOP_{extra}$) count associated with the presence of immersed boundaries versus the case with simple domain ($N$ grid points) showed that $FLOP_{extra}\approx O(N_{irr}/N)$, where $N_{irr}$ is the number of irregular grid points. Two factors contribute to the cost-effectiveness and efficiency of the new method: (1) the preconditioner cost which is independent of the presence of an immersed boundary, and (2) the compact nature of the finite-difference discretization, which minimizes the additional operations per irregular grid point.

The accuracy and the efficiency of the new method was demonstrated through verification using problems with analytical solutions for domains with and without immersed boundaries on uniform and nonuniform grids. The numerical results demonstrate that the solutions are sharp across the interface and that the new method is fourth-order-accurate in the maximum norm, including the irregular grid points near the interface of the immersed boundary. For the test cases investigated, the solution technique was found to be very efficient in the sense that the accuracy and the convergence behavior do not deteriorate with the presence of irregular boundaries while the $FLOP_{extra}$ per outer iteration remains insignificant.

\addcontentsline{toc}{chapter}{Acknowledgements}
\section*{Acknowledgements} \label{ack} 
This work was supported in part by the Air Force Office of Scientific Research (AFOSR) under grant number FA9550-14-1-0184 and by National Science Foundation (NSF) under grant number
1805273, with Dr D. Smith and Dr R. Joslin serving as the program manager, respectively.

\addcontentsline{toc}{chapter}{References}
 \bibliographystyle{elsarticle-num-names}

\bibliography{JCP}

\begin{thebibliography}{27}
\providecommand{\natexlab}[1]{#1}
\providecommand{\url}[1]{\texttt{#1}}
\providecommand{\urlprefix}{URL }
\expandafter\ifx\csname urlstyle\endcsname\relax
  \providecommand{\doi}[1]{doi:\discretionary{}{}{}#1}\else
  \providecommand{\doi}[1]{doi:\discretionary{}{}{}\begingroup
  \urlstyle{rm}\url{#1}\endgroup}\fi
\providecommand{\bibinfo}[2]{#2}

\bibitem[{Leveque and Li(1994)}]{17}
\bibinfo{author}{R.~J. Leveque}, \bibinfo{author}{Z.~Li}, \bibinfo{title}{The
  immersed interface method for elliptic equations with discontinuous
  coefficients and singular sources}, \bibinfo{journal}{SIAM J. Numer. Anal.}
  \bibinfo{volume}{31} (\bibinfo{year}{1994}) \bibinfo{pages}{1019--1044}.

\bibitem[{Spotz and Carey(1996)}]{1}
\bibinfo{author}{W.~F. Spotz}, \bibinfo{author}{G.~F. Carey}, \bibinfo{title}{A
  high-order compact formulation for the 3D Poisson equation},
  \bibinfo{journal}{Num. Meth. for Part. Dif. Eq.} \bibinfo{volume}{12}
  (\bibinfo{year}{1996}) \bibinfo{pages}{235--243}.

\bibitem[{Wang et~al.(2006)Wang, Zhong, and Zhang}]{2}
\bibinfo{author}{J.~Wang}, \bibinfo{author}{W.~Zhong},
  \bibinfo{author}{J.~Zhang}, \bibinfo{title}{A general meshsize fourth-order
  compact difference discretization scheme for 3D Poisson equation},
  \bibinfo{journal}{Applied Math. and Comp.} \bibinfo{volume}{183}
  (\bibinfo{year}{2006}) \bibinfo{pages}{804--812}.

\bibitem[{Sutmann and Steffen(2006{\natexlab{a}})}]{3}
\bibinfo{author}{G.~Sutmann}, \bibinfo{author}{B.~Steffen},
  \bibinfo{title}{High-order compact solvers for the three-dimensional Poisson
  equation}, \bibinfo{journal}{J. Comput. Appl. Math.} \bibinfo{volume}{187}
  (\bibinfo{year}{2006}{\natexlab{a}}) \bibinfo{pages}{142--170}.

\bibitem[{Ge(2010)}]{4}
\bibinfo{author}{Y.~Ge}, \bibinfo{title}{Multigrid method and fourth-order
  compact difference discretization scheme with unequal meshsizes for 3D
  Poisson equation}, \bibinfo{journal}{J. Comp. Phys.} \bibinfo{volume}{229}
  (\bibinfo{year}{2010}) \bibinfo{pages}{6381–6391}.

\bibitem[{Ge et~al.(2013)Ge, Cao, and Zhang}]{5}
\bibinfo{author}{Y.~Ge}, \bibinfo{author}{F.~Cao}, \bibinfo{author}{J.~Zhang},
  \bibinfo{title}{A transformation-free HOC scheme and multigrid method for
  solving the 3D Poisson equation on nonuniform grids}, \bibinfo{journal}{J.
  Comp. Phys.} \bibinfo{volume}{234} (\bibinfo{year}{2013})
  \bibinfo{pages}{199--216}.

\bibitem[{Nejat and Ollivier-Gooch(2003)}]{6}
\bibinfo{author}{A.~Nejat}, \bibinfo{author}{C.~Ollivier-Gooch},
  \bibinfo{title}{A high-order accurate Unstructured GMRES solver for Poisson`s
  equation}, in: \bibinfo{booktitle}{11th annual conference of the
  computational fluid dynamics society of Canada}, \bibinfo{year}{2003}.

\bibitem[{Peskin(1972)}]{7}
\bibinfo{author}{C.~S. Peskin}, \bibinfo{title}{Flow patterns around heart
  valves: a numerical method}, \bibinfo{journal}{J. Comput. Phys.}
  (\bibinfo{year}{1972}) \bibinfo{pages}{252--271}.

\bibitem[{Peskin(1977)}]{8}
\bibinfo{author}{C.~S. Peskin}, \bibinfo{title}{Numerical analysis of blood
  flow in the heart}, \bibinfo{journal}{J. Comput. Phys.} \bibinfo{volume}{25}
  (\bibinfo{year}{1977}) \bibinfo{pages}{220--252}.

\bibitem[{Mittal and Iaccarino(2005)}]{9}
\bibinfo{author}{R.~Mittal}, \bibinfo{author}{G.~Iaccarino},
  \bibinfo{title}{Immersed boundary methods}, \bibinfo{journal}{Annu. Rev.
  Fluid Mech.} \bibinfo{volume}{37} (\bibinfo{year}{2005})
  \bibinfo{pages}{239--261}.

\bibitem[{Seo and Mittal(2011)}]{10}
\bibinfo{author}{J.~H. Seo}, \bibinfo{author}{R.~Mittal}, \bibinfo{title}{A
  high-order immersed boundary method for acoustic wave scattering and low-Mach
  number flow-induced sound in complex geometries}, \bibinfo{journal}{J. Comp.
  Physics} \bibinfo{volume}{230}~(\bibinfo{number}{4}) (\bibinfo{year}{2011})
  \bibinfo{pages}{1000 -- 1019}.

\bibitem[{Tseng and Ferziger(2003)}]{11}
\bibinfo{author}{Y.~H. Tseng}, \bibinfo{author}{J.~H.~. Ferziger},
  \bibinfo{title}{A ghost-cell immersed boundary method for flow in complex
  geometry}, \bibinfo{journal}{J. Comput. Phys.} \bibinfo{volume}{192}
  (\bibinfo{year}{2003}) \bibinfo{pages}{593--623}.

\bibitem[{Mittal et~al.(2008)Mittal, Dong, Bozkurttas, Najjar, Vargas, and von
  Loebbecke}]{12}
\bibinfo{author}{R.~Mittal}, \bibinfo{author}{H.~Dong},
  \bibinfo{author}{M.~Bozkurttas}, \bibinfo{author}{F.~M. Najjar},
  \bibinfo{author}{A.~Vargas}, \bibinfo{author}{A.~von Loebbecke},
  \bibinfo{title}{A versatile sharp interface immersed boundary method for
  incompressible flows with complex boundaries}, \bibinfo{journal}{J. Comp.
  Phys.} \bibinfo{volume}{227} (\bibinfo{year}{2008})
  \bibinfo{pages}{4825--4852}.

\bibitem[{Zhu et~al.(2017)Zhu, Seo, Vedula, and R.}]{13}
\bibinfo{author}{C.~Zhu}, \bibinfo{author}{J.~H. Seo},
  \bibinfo{author}{V.~Vedula}, \bibinfo{author}{M.~R.}, \bibinfo{title}{A
  highly scalable sharp-interface immersed boundary method for large-scale
  parallel computers}, in: \bibinfo{booktitle}{AIAA Paper},
  \bibinfo{publisher}{AIAA 2017-3622}, \bibinfo{year}{2017}.

\bibitem[{Udaykumar et~al.(2001)Udaykumar, Mittal, Rampunggoon, and
  Khanna}]{14}
\bibinfo{author}{H.~Udaykumar}, \bibinfo{author}{R.~Mittal},
  \bibinfo{author}{P.~Rampunggoon}, \bibinfo{author}{A.~Khanna},
  \bibinfo{title}{A sharp interface cartesian grid method for simulating flows
  with complex moving boundaries}, \bibinfo{journal}{J. Comp. Physics}
  \bibinfo{volume}{174}~(\bibinfo{number}{1}) (\bibinfo{year}{2001})
  \bibinfo{pages}{345 -- 380}.

\bibitem[{Udaykumar et~al.(2002)Udaykumar, Mittal, and Rampunggoon}]{15}
\bibinfo{author}{H.~S. Udaykumar}, \bibinfo{author}{R.~Mittal},
  \bibinfo{author}{P.~Rampunggoon}, \bibinfo{title}{Interface tracking finite
  volume method for complex solid–fluid interactions on fixed meshes},
  \bibinfo{journal}{Communications in Numerical Methods in Engineering}
  \bibinfo{volume}{18}~(\bibinfo{number}{2}) (\bibinfo{year}{2002})
  \bibinfo{pages}{89--97}.

\bibitem[{S. and Mittal(2011)}]{16}
\bibinfo{author}{J.~H. S.}, \bibinfo{author}{R.~Mittal}, \bibinfo{title}{A
  sharp-interface immersed boundary method with improved mass conservation and
  reduced spurious pressure oscillations}, \bibinfo{journal}{J. Comp. Physics}
  \bibinfo{volume}{230}~(\bibinfo{number}{19}) (\bibinfo{year}{2011})
  \bibinfo{pages}{7347 -- 7363}.

\bibitem[{Linnick and Fasel(2004)}]{18}
\bibinfo{author}{M.~Linnick}, \bibinfo{author}{H.~F. Fasel}, \bibinfo{title}{A
  high-order immersed interface method for simulating unsteady incompressible
  flows on irregular domains}, \bibinfo{journal}{J. Comput. Phys.}
  \bibinfo{volume}{204} (\bibinfo{year}{2004}) \bibinfo{pages}{157--192}.

\bibitem[{Hosseinverdi and Fasel(2017)}]{20}
\bibinfo{author}{S.~Hosseinverdi}, \bibinfo{author}{H.~F. Fasel},
  \bibinfo{title}{Very high-order accurate sharp immersed interface method:
  application to direct numerical simulations of incompressible flows}, in:
  \bibinfo{booktitle}{23rd AIAA Computational Fluid Dynamics Conference},
  \bibinfo{publisher}{AIAA 2017-3624}, \bibinfo{year}{2017}.

\bibitem[{Jomaa and Macaskill(2010)}]{21}
\bibinfo{author}{Z.~Jomaa}, \bibinfo{author}{C.~Macaskill}, \bibinfo{title}{The
  Shortley–Weller embedded finite-difference method for the 3D Poisson
  equation with mixed boundary conditions}, \bibinfo{journal}{J. Comp. Physics}
  \bibinfo{volume}{229}~(\bibinfo{number}{10}) (\bibinfo{year}{2010})
  \bibinfo{pages}{3675 -- 3690}.

\bibitem[{Yu and Wei(2007)}]{22}
\bibinfo{author}{S.~Yu}, \bibinfo{author}{G.~Wei},
  \bibinfo{title}{Three-dimensional matched interface and boundary (MIB) method
  for treating geometric singularities}, \bibinfo{journal}{J. Comp. Physics}
  \bibinfo{volume}{227}~(\bibinfo{number}{1}) (\bibinfo{year}{2007})
  \bibinfo{pages}{602 -- 632}.

\bibitem[{Hosseinverdi and Fasel(2018)}]{HOSSEINVERDI2018912}
\bibinfo{author}{S.~Hosseinverdi}, \bibinfo{author}{H.~F. Fasel},
  \bibinfo{title}{An efficient, high-order method for solving Poisson equation
  for immersed boundaries: Combination of compact difference and multiscale
  multigrid methods}, \bibinfo{journal}{J. Comp. Phys.} \bibinfo{volume}{374}
  (\bibinfo{year}{2018}) \bibinfo{pages}{912 -- 940}.

\bibitem[{Sutmann and Steffen(2006{\natexlab{b}})}]{steffen}
\bibinfo{author}{G.~Sutmann}, \bibinfo{author}{B.~Steffen},
  \bibinfo{title}{High-order compact solvers for the three-dimensional Poisson
  equation}, \bibinfo{journal}{J. Comp. and App. Math.}
  \bibinfo{volume}{187}~(\bibinfo{number}{2})
  (\bibinfo{year}{2006}{\natexlab{b}}) \bibinfo{pages}{142 -- 170}, ISSN
  \bibinfo{issn}{0377-0427}.

\bibitem[{Sutmann(2007)}]{sutmann}
\bibinfo{author}{G.~Sutmann}, \bibinfo{title}{Compact finite difference schemes
  of sixth order for the Helmholtz equation}, \bibinfo{journal}{J. Comp. and
  App. Math.} \bibinfo{volume}{203}~(\bibinfo{number}{1})
  (\bibinfo{year}{2007}) \bibinfo{pages}{15 -- 31}, ISSN
  \bibinfo{issn}{0377-0427}.

\bibitem[{Zhong(2007)}]{ZHONG2007}
\bibinfo{author}{X.~Zhong}, \bibinfo{title}{A new high-order immersed interface
  method for solving elliptic equations with imbedded interface of
  discontinuity}, \bibinfo{journal}{Journal of Computational Physics}
  \bibinfo{volume}{225}~(\bibinfo{number}{1}) (\bibinfo{year}{2007})
  \bibinfo{pages}{1066 -- 1099}.

\bibitem[{Saad(1996)}]{23}
\bibinfo{author}{Y.~Saad}, \bibinfo{title}{Iterative Methods for Sparse Linear
  Systems}, \bibinfo{publisher}{PWS Publishing}, \bibinfo{address}{New York},
  \bibinfo{year}{1996}.

\bibitem[{Zedan and Schneider(1983)}]{24}
\bibinfo{author}{M.~Zedan}, \bibinfo{author}{G.~E. Schneider},
  \bibinfo{title}{A three-dimensional modified strongly implicit procedure for
  heat conduction}, \bibinfo{journal}{AIAA Journal} \bibinfo{volume}{21}
  (\bibinfo{year}{1983}) \bibinfo{pages}{295--303}.

\end{thebibliography}

\end{document}